\documentclass[3p]{elsarticle}
\usepackage{pstricks}
\usepackage{pst-eps}
\usepackage{natbib}

\usepackage{amsthm,amsmath,mathrsfs,amssymb}
\usepackage{epstopdf}
\usepackage{bbm}
\usepackage{color}
\usepackage{comment}
\usepackage{multirow}

\usepackage{float}
\usepackage{stfloats}
\usepackage[tight]{subfigure}
\usepackage{graphicx,color,url}

\usepackage{keyval}
\usepackage{algorithm}
\usepackage{algorithmic}
\usepackage{tabularx,booktabs}
\usepackage[colorlinks, citecolor=blue]{hyperref}
\usepackage{extarrows}
\usepackage{stackengine}
\usepackage{bm}

\DeclareMathOperator{\rank}{rank}
\DeclareMathOperator{\spn}{span}

\newsavebox{\tempbox}
\newtheorem{definition}{Definition}

\newtheorem{theorem}{Theorem}
\newtheorem{remark}{Remark}
\begin{document}

\begin{frontmatter}

\title{A Low-rank Spline Approximation of Planar Domains}

\author{Maodong Pan}
\ead{mdpan@mail.ustc.edu.cn}
\author{Falai Chen\corref{cor}}
\ead{chenfl@ustc.edu.cn}
\cortext[cor]{Corresponding author.}

\address{School of Mathematical Sciences, University of Science and Technology of China, Hefei, Anhui, 230026, PR China}

\begin{abstract}
Construction of spline surfaces from given boundary curves is one of the classical problems in computer aided geometric design, which regains much attention in isogeometric analysis in recent years and is called domain parameterization. However, for most of the state-of-the-art parameterization methods, the rank of the spline parameterization is usually large, which results in higher computational cost in solving numerical PDEs. In this paper, we propose a low-rank representation for the spline parameterization of planar domains using low-rank tensor approximation technique, and apply quasi-conformal map as the framework of the spline parameterization. Under given correspondence of boundary curves, a quasi-conformal map with low rank and low distortion between a unit square and the computational domain can be modeled as a non-linear optimization problem. We propose an efficient algorithm to compute the quasi-conformal map by solving two convex optimization problems alternatively. Experimental results show that our approach can produce a bijective and low-rank parametric spline representation of planar domains, which results in better performance than previous approaches in solving numerical PDEs.
\end{abstract}


\end{frontmatter}
% main text

%!Mode:: "TeX:UTF-8"

%!Mode:: "TeX:UTF-8"

\section{Introduction}
\label{sec:intro}
Given the boundary curves in 2D or 3D, constructing a parametric spline representation to interpolate the given boundary
is a fundamental problem in Computer Aided Geometric Design, and Coons surfaces are a classic tool to solve the problem~\citep{farin1999discrete}. This problem has been revived in recent years due to its applications in isogeometric analysis (IGA), and it is called $\emph{domain parameterization}$.  Domain parameterization has a great effect on the accuracy and efficiency in subsequent analysis~\citep{cohen2010analysis,xu2011parameterization,pilgerstorfer2014bounding}. It is a common requirement that
the parameterization should be injective, i.e., the mapping from the parametric domain (generally a unit square) to the computational domain is self-intersection free. In addition, the distortion of the map should be as small as possible, i.e.,
the areas and angles after mapping should be preserved as much as possible. So far many approaches have been proposed to solve the parameterization problem, e.g. the discrete Coons interpolation~\citep{farin1999discrete}, the harmonic mapping~\citep{martin2009volumetric,nguyen2010parameterization,xu2011variational}, the spring model~\citep{gravesen2012planar}, the nonlinear optimization method~\citep{xu2011parameterization}, parameterization with non-standard B-splines (e.g., T-splines~\citep{zhang2012solid,zhang2013conformal}, THB-splines~\citep{falini2015planar}), the method based on Teichm{\"u}ller mapping~\citep{nian2016planar} and so on. While all these methods focus on low distortion and  bijectivity of the parameterization, the problem of low-rank parameterization is not discussed. In fact, the rank of the parametric spline representation by these methods is usually large, which results in higher computational cost in subsequent isogeometric analysis~\citep{mantzaflaris2014matrix,juttler2017low}. Recently, Juetter
and his collaborators have observed that reducing the rank of a parameterization can lead to substantial improvements of the overall efficiency of the numerical simulation~\citep{mantzaflaris2014matrix,mantzaflaris2017low,juttler2017low}. This observation motivates us to explore parameterization techniques which are able to generate low-rank spline representations.

In this paper, by using low-rank tensor approximation technique, we propose a low-rank representation for planar domain parameterization based on quasi-conformal mapping. Quasi-conformal mapping is a natural extension of conformal mapping which preserves angles~\citep{lehto1973quasiconformal,lui2010compression}. The angular distortion and the bijectivitiy of a quasi-conformal map can be characterized by a complex function called the \emph{Beltrami coefficient}~\citep{lui2012optimization,lui2013texture}. By optimizing the norm of the Beltrami coefficient
and the rank of the spline representation, we are able to find a planar domain parameterization with low rank and low distortion as much as possible.

%In mathematics, the theory of quasi-conformal mapping has a long history and~\citep{lehto1973quasiconformal} introduces it in detail. Quasi-conformal mapping is a natural extension of conformal mapping with bounded conformal distortion. The nonconformality of a quasi-conformal mapping is characterized by a complex function called \emph{Beltrami coefficient}. The norm of the Beltrami coefficient measures the conformal distortion of the map. The theory and applications of quasi-conformal mapping are discussed in~\citep{zeng2009surface,zeng2011registration,zeng2012computing,lui2012optimization,lam2014landmark,lui2015splitting,lee2016landmark,nian2016planar}, where the problems of surface and image registration, parameterization, etc. are considered.

The remainder of this paper is organized as follows. Section~\ref{sec:work} reviews some related work about domain parameterization and the applications of low-rank tensors in science and engineering. Section~\ref{sec:preliminaries} presents some preliminary knowledge about quasi-conformal mapping and low-rank tensor approximation. In Section~\ref{sec:approach}, we propose a mathematical model followed by an algorithm to compute a low-rank quasi-conformal map for domain parameterization. Section~\ref{sec:result} demonstrates some experimental results of our algorithm and its applications in solving numerical PDEs. Comparisons with the state-of-the-art methods are also provided. Finally, we conclude the paper with a summary and future work in Section~\ref{sec:conclusion}.

\section{Related work}
\label{sec:work}
\subsection{Domain parameterization}
Domain parameterization is one essential step in isogeometric analysis~\citep{hughes2005isogeometric}.
The quality of the parameterization greatly influences the numerical accuracy and efficiency of the numerical simulations~\citep{cohen2010analysis,xu2013optimal,pilgerstorfer2014bounding}. Over the past decade, many approaches have been proposed to solve the parameterization problem. A simple way for domain parameterization is based on discrete Coons patches proposed by Farin and Hansford~\citep{farin1999discrete}. A spring model was suggested by Gravesen et al. to solve the problem
~\citep{gravesen2012planar}. The harmonic functions have many good properties and they were used in~\citep{martin2009volumetric,nguyen2010parameterization,xu2011variational} to construct domain parameterizations.
These methods are generally computational inexpensive but the resulting parameterization may not be injective---a deficiency that should be avoided in such type of applications.
Xu et al.~\citep{xu2011parameterization} presented a sophisticated nonlinear optimization technique with the injectivity and the quality of the parameterization as an objective. In~\citep{falini2015planar}, THB-splines is used for planar domain parameterization with varying levels of computational complexity. Recently, Nian et al.~\citep{nian2016planar} proposed an approach for planar domain parameterization based on Teichm\"{u}ller mapping, which guarantees a bijective and high-quality parameterization. For 3D domains,
a framework was developed in~\citep{martin2009volumetric} to model a single trivariate B-spline from input boundary triangle meshes with genus-zero topology. Aigner et al.~\citep{aigner2009swept} presented a variational framework for generating NURBS parametrizations of swept volumes, in which the control points can be obtained by solving an optimization problem. Escobar et al.~\citep{escobar2011new} proposed a solid T-spline modeling algorithm from a surface triangular mesh. Zhang et al.~\citep{zhang2012solid} developed a mapping-based method to construct rational trivariate solid T-splines for genus-zero geometry from the boundary triangulations. For meshes with more general geometry, they~\citep{wang2013trivariate} further used the mapping, subdivision and pillowing techniques to generate high quality T-spline representations. In~\citep{xu2013constructing}, the authors proposed a variational harmonic method to construct analysis-suitable parameterization of a computational domain from given CAD boundary information. For models topologically equivalent to a set of cubes and bounded by B-spline surfaces, they~\citep{xu2013analysis} further studied the volume parameterization of the multi-block computational domain using the nonlinear optimization method proposed in~\citep{xu2011parameterization}. When dealing with more complex geometric shapes, however, single-patch representations do not provide sufficient flexibility. Multi-patch structures are generally constructed to fulfill the task of low distortion parameterization~\citep{xu2015two,buchegger2017planar}. In this paper, we focus on 2D domain parameterization.

%While all the above methods focus on the injectivity and quality of the parameterization, very few work has concerned with the rank of the parameterization which has much impact on the computational efficiency in subsequent analysis.

\subsection{Applications of low-rank tensor approximation}
Low-rank approximation is very helpful for dimension reduction and data compression, and has been successfully applied in many fields like signal processing, computer vision, patter recognition, computer graphics, etc. A thorough survey on this topic is out of the scope of this paper, and we refer the reader to~\citep{markovsky2011low,ma2012sparse} and references therein. Tensors, as a generation of matrices in higher dimensions have important applications  in science and engineering, e.g., psychometrics, psychometrics and data mining~\citep{kolda2009tensor}. The details of low-rank tensor approximation and its applications have been discussed in depth in~\citep{grasedyck2013literature}. Recently, the low-rank tensor optimization has been applied in graphics and geometric modeling community, e.g., in finding the upright orientation of 3D shapes~\citep{wang2014upright} and
in compact implicit surface reconstructions~\citep{pan2016compact}. For other applications of low-rank tensors in geometric modeling and processing, please refer to~\citep{xu2015survey} and references therein.
Juetter and his collaborators recently addressed the problem of low-rank approximation for isogeometric analysis applications. Mantzaflaris et al.~\citep{mantzaflaris2014matrix} applied low-rank matrix approximation for accelerating the assembly process of stiffness matrices in isogeometric analysis. They further extended their work to 3D case and employed the tensor decomposition technique for Galerkin-based isogeometric analysis, which can reduce the computation time and storage requirements dramatically~\citep{mantzaflaris2017low}. A construction for low-rank tensor-product spline surfaces from given boundary curves is also proposed by J\"{u}ttler et al.~\citep{juttler2017low}.

\section{Preliminaries}
\label{sec:preliminaries}
In this section, we give some preliminary knowledge about quasi-conformal mapping and low-rank tensor approximation followed by the definition of rank-$R$ spline functions.
\subsection{Quasi-conformal mapping}
\label{sec:QCMap}
The most convenient way to explain quasi-conformal mapping is in complex setting. Let $z=x+iy$ be a complex variable with $x$ and $y$ being the real and imaginary part of $z$ respectively, and $\bar{z}=x-iy$ be the conjugate of $z$,
here $i=\sqrt{-1}$. For a differentiable complex function $f(z)$, its complex derivatives are defined as $f_{z}=\frac{1}{2}(f_x-if_y)$ and $f_{\bar{z}}=\frac{1}{2}(f_x+if_y)$. A complex function defines a map from a complex plane
to a complex plane. When $f_{\bar{z}}=0$, $f$ defines a \emph{conformal map} which preserves angles and maps an infinitesimal circle to an infinitesimal circle. A quasi-conformal map is a generalization of a conformal map which
maps an infinitesimal circle to an infinitesimal ellipse.

\begin{definition}\label{QCDefinition}
Suppose $f:\hat{\Omega}\to \Omega$ is a complex function, where $\hat{\Omega}$ and $\Omega$ are two domains in $\mathbb{C}$. If $f$ is assumed to have continuous partial derivatives, then $f$ is quasi-conformal provided it satisfies the Beltrami equation
\begin{equation}\label{QuasiConformal}
\frac{\partial f}{\partial \bar{z}}=\mu(z)\frac{\partial f}{\partial z}
\end{equation}
for some complex valued Lebesgue measurable $\mu$ satisfying $\sup|\mu|<1$. $\mu$ is called the Beltrami coefficient of the map $f$.
\end{definition}

\begin{figure}[!htbp]
    \centering
        \includegraphics[width=0.40\textwidth]{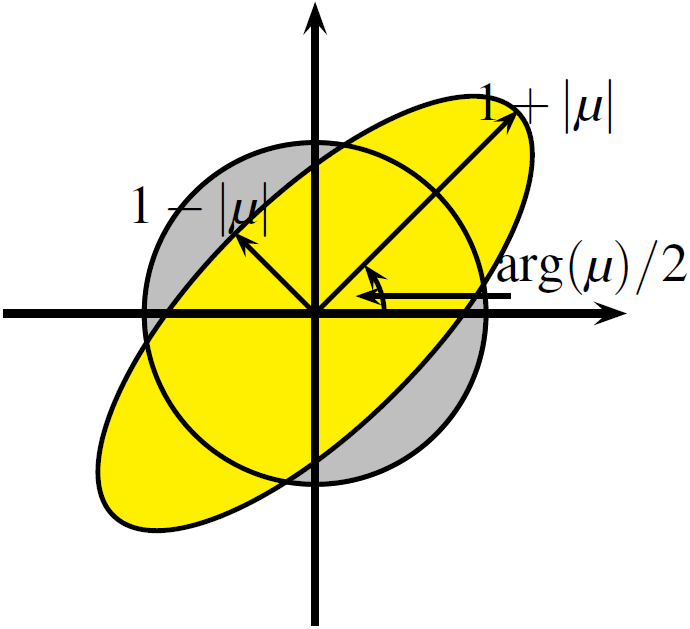}
     \caption{Geometric meaning of the Beltrami coefficient.}
    \label{BeltramiCoefficient}
\end{figure}

The Beltrami coefficient $\mu$ determines the angular deviation from conformality. When $\mu=0$, the quasi-conformal map becomes conformal. Define the dilatation of $f$ at the point $z$ by
$$K(z)=\frac{1+|\mu(z)|}{1-|\mu(z)|}.$$
Then a quasi-conformal map takes infinitesimal circles to infinitesimal ellipses with bounded eccentricity given by the dilatation $K=\frac{1+\|\mu\|_{\infty}}{1-\|\mu\|_{\infty}}$ and the orientation of axis rotates an
$\arg(\mu)/2$, as shown in Fig.~\ref{BeltramiCoefficient}. Furthermore, $f(z)$ is orientation preserving and bijective provided $\|\mu\|_{\infty} < 1$ and $f_z\neq0$.

Besides angular deviation, another quantity that characterizes the area distortion of a map $f$ is the Jacobian $J(f)$ of the map. In order to eliminate the area difference between the parametric domain (a unit square) and the computational domain, usually scaled Jacobin $J_s(f)$ is employed:
\begin{equation*}
     J_s(f)=\frac{J(f)}{A_{\Omega}}
\end{equation*}
where $A_{\Omega}$ is the area of the computational domain $\Omega$.
%So the area distortion for a subdomain $\mathbf{d}$ of the parametric domain $\hat{\Omega}$, denoted as $\delta^{area}(\mathbf{d})$, can be calculated as follows:
%\begin{equation}\label{areadistortion}
%\delta^{area}(\mathbf{d}) = \frac{A_{\mathbf{d}}}{\int_{\mathbf{d}}J_s(f)\mathrm{d}z}
%\end{equation}
%where $A_{\mathbf{d}}$ is the area of $\mathbf{d}$.
%\begin{theorem}(\citep{ahlfors1966lectures})\label{QCExist}
%If $f:D\to D'$ satisfies $\|\mu(f)\|_{\infty}<1$, then $f$ is a diffeomorphism.
%\end{theorem}
%Theorem~\ref{QCExist} plays the fundamental role of obtaining a diffeomorphism. That is, given a $\mu$ with $\|\mu\|_{\infty}<1$, there is a quasi-conformal mapping from $D$ to $D'$ satisfying the Beltrami equation~(\ref{QCDefinition}).

\subsection{Low-rank tensor approximation}
\label{sec:LRTA}

A \textit{tensor} is a multidimensional array. More formally, an $n$th-order or $n$-way tensor is an element of the tensor product of $n$ vector spaces, each of which has its own coordinate system \citep{kolda2009tensor}. An $n$th-order tensor is usually denoted by boldface Euler script letters, e.g., $\mathcal{X}\in\mathbb{C}^{I_{1}\times I_{2}\times \ldots I_{n}}$. A first-order tensor is a vector, a second-order tensor is a matrix, and tensors of order three or higher are called higher-order tensors.

An $n$th-order tensor $\mathcal{X}\in\mathbb{C}^{I_{1}\times I_{2}\times \ldots I_{n}}$ is \textit{rank one} if it can be written as the outer product of $n$ vectors, i.e.,
\begin{equation*}\label{RankOne}
    \mathcal{X} = \mathbf{a}^{(1)}\circ\mathbf{a}^{(2)}\circ\cdots\circ\mathbf{a}^{(n)},
\end{equation*}
where $\circ$ denotes the outer product, and $\mathbf{a}^{(i)}\in\mathbb{C}^{I_i}$.\\

\noindent\textbf{CP decomposition}
Let $\mathcal{X}\in\mathbb{C}^{I_{1}\times I_{2}\times \ldots I_{n}}$ be an $n$th-order tensor, the \textit{CP decomposition} factorizes $\mathcal{X}$ into a sum of component rank-one tensors as follows:
\begin{equation}\label{CPDDefi}
    \mathcal{X} = \sum_{r=1}^{R}\lambda_r\mathbf{a}_{r}^{(1)}\circ\mathbf{a}_{r}^{(2)}\circ\cdots\circ\mathbf{a}_{r}^{(n)},
\end{equation}
where $R$ is a positive integer and $\lambda_r>0, \ \mathbf{a}_{r}^{(i)}\in\mathbb{C}^{I_i}$ for $r = 1,\ldots,R$. It's often useful to assume that $\mathbf{a}_{r}^{(i)}$ are normalized to length one with the weights absorbed into $\lambda_{r}$. \\

The \textit{rank} of a tensor $\mathcal{X}$, denoted as $\rank(\mathcal{X})$, is the smallest number of components in the above expression~(\ref{CPDDefi}). The CP decomposition can be considered to be a higher-order generalization of the matrix singular value decomposition (SVD) which can be described as follows:\\

\noindent\textbf{Singular value decomposition}
Let $X\in\mathbb{C}^{I_{1}\times I_{2}}$ be a matrix, the \textit{SVD} factorizes $X$ into a sum of component rank-one matrices as follows:
\begin{equation}\label{SVDDefi}
    X = U\Sigma V^\mathrm{H} = \sum_{r=1}^{R}\lambda_r\mathbf{a}_{r}\circ\mathbf{b}_{r}
\end{equation}
where $U$ is an $I_{1}\times I_{1}$ complex unitary matrix, $\Sigma$ is an $I_{1}\times I_{2}$ rectangular diagonal matrix with non-negative real numbers
on the diagonal, and $V$ is an $I_{2}\times I_{2}$ complex unitary matrix. The diagonal entries of $\Sigma$ are known as the singular values of $X$ and $\lambda_{r}$ are the nonzero singular values. $\mathbf{a}_{r}$, $\mathbf{b}_{r}$ are the left $R$ column vectors of $U$ and $V$ respectively.

\begin{theorem}(\citep{eckart1936approximation})\label{bestapproximation}
The best rank-$k$ approximation of $X$ is given by a truncated SVD of $X$, that is
\begin{equation}\label{LRA}
\hat{X} = U\hat{\Sigma}V^\mathrm{H}=\sum_{r=1}^{k}\lambda_r\mathbf{a}_{r}\circ\mathbf{b}_{r}
\end{equation}
\end{theorem}
where $\hat{X}$ has a specific rank $k$, and $\hat{\Sigma}$ is the same matrix as $\Sigma$ except that it contains only the $k$ largest singular values (the other singular values are replaced by zero). $\hat{X}$ is called rank-$k$ approximation of $X$.

From Theorem~\ref{bestapproximation}, we can see that the \textit{rank} of a matrix $X$, denoted as $\rank(X)$, is equal to the number of nonzero singular values in SVD of $X$. However, $\rank(X)$ is a nonconvex function, and solving a rank-constrained problem is generally NP-hard.
Recently several works \citep{recht2010guaranteed,cai2010singular,liu2013tensor} use the trace norm of a matrix to approximately calculate the rank, which leads to a convex optimization problem.
The trace norm of $X$ is defined as follows
\begin{equation}\label{nuclearnorm}
    \|X\|_{\ast}:=\sum_{i}\sigma_{i}(X)
\end{equation}
where $\sigma_{i}(X)$ is the $i$th largest singular value of $X$.

\subsection{Rank-$R$ spline functions}
\label{sec:rankRfunction}

A multivariate function $f(x_1,\ldots,x_n)$ is said to have {\em rank} $R$  if it can be represented as a sum of separable functions
\begin{equation}\label{rank-Rf}
f(x_1,\ldots,x_n)=\sum_{r=1}^{R}\prod_{k=1}^{n}f_{r}^{(k)}(x_k)
\end{equation}
where $f_{r}^{(k)}(x_k)$ are univariate functions.

Let $g(x_1,\ldots,x_n)$ be an $n$-variate tensor product spline function of $n$-degree ($d_1,\ldots,d_n$) defined over an $n$-dimensional domain $\Omega$:
\begin{equation}\label{Alegraic_spline}
g(x_1,\ldots,x_n) = \sum_{i_1,i_2,\ldots,i_n}c_{i_1,i_2,\ldots,i_n}\prod_{k=1}^{n}\beta_{i_k}^{(k)}(x_k)
\end{equation}
where $\left\{c_{i_1,i_2,\ldots,i_n}\right\}$ are the control coefficients, $\left\{\beta_{i_k}^{(k)}(x_k)\right\}$ are B-spline basis functions in $n$ possibly different univariate spline spaces $\mathcal{S}_{\boldsymbol{\tau}_{k}}^{d_k}, k=1,\ldots,n$, and each space is defined by a knot vector $\boldsymbol{\tau}_{k}$ and a degree $d_k$. We collect the basis functions in the knot vector $\boldsymbol{\tau}_{k}$
\begin{equation*}
\boldsymbol{\beta}^{(k)}(x_k)=\left[\beta_{i_k}^{(k)}(x_k)\right],\quad k=1,\ldots,n,
\end{equation*}
Let $\mathcal{C}$ be the $n$-order coefficient tensor associated with the coefficients $\{c_{i_1,i_2,\ldots,i_n}\}$ defined in~(\ref{Alegraic_spline}). If the rank of $\mathcal{C}$ is R and perform the CP decomposition of $\mathcal{C}$ as
\begin{equation}
\mathcal{C}\overset{(\ref{CPDDefi})}=\sum_{r=1}^{R}\mathbf{c}_{r}^{(1)}\circ\mathbf{c}_{r}^{(2)}\circ\cdots\circ\mathbf{c}_{r}^{(n)},
\end{equation}
then $g$ can be expressed in a sum of $R$ products
\begin{equation}\label{rank-Rsplinef}
g(x_1,\ldots,x_n)=\sum_{r=1}^{R}\prod_{k=1}^{n}g_{r}^{(k)}(x_k)
\end{equation}
of the univariate spline functions
\begin{equation*}
g_{r}^{(k)}(x_k)=\mathbf{c}_{r}^{(k)}\cdot\boldsymbol{\beta}^{(k)}(x_k).
\end{equation*}
Thus $g$ also has rank $R$, and we call $g$ is a \textit{rank-$R$ spline function}.

\section{Parameterization of computational domains via low-rank tensor approximation}
\label{sec:approach}

\subsection{Representation of parameterization}
\begin{figure}[!htbp]
    \centering
        \includegraphics[width=0.50\textwidth]{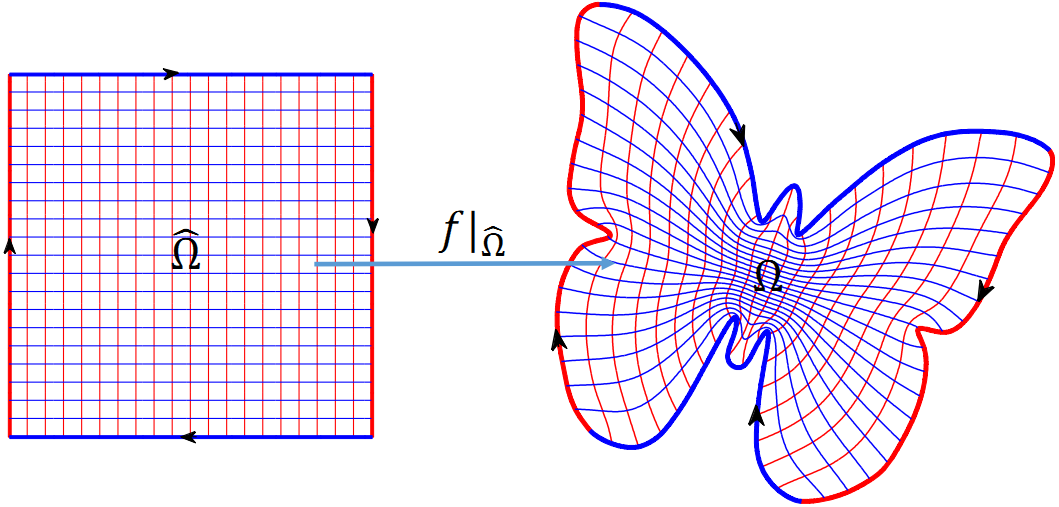}
     \caption{Parameterization--a map $f$ from a unit square $\hat{\Omega}$ to a computational domain $\Omega$.}
    \label{BoundaryMapping}
\end{figure}

Suppose we are given the B-spline representations of the four boundary curves of a computational domain $\Omega$. Our aim is to compute a B-spline representation for the parameterization domain $\Omega$, that is, a map from the unit square $\hat{\Omega}=[0,1]^2$ to
$\Omega$ which is bijective, low distortion and low rank. An example is illustrated in Fig.~\ref{BoundaryMapping}.

Assume the parameterization of the computation domain $\Omega$ is expressed by a tensor product B-spline function
\begin{equation}\label{B-spline rep}
    \begin{split}
{\bf P}(x,y):=\sum_{i=0}^{m}\sum_{j=0}^{n}\bm{P}_{ij}M_{i}^{p}(x)N_{j}^{q}(y)
    \end{split}
\end{equation}
where $\bm{P}_{ij}=(x_{ij},y_{ij})\in \mathbb{R}^2$ are the control points, $M_{i}^{p}(x)$ and $N_{j}^{q}(y)$ are the B-spline basis functions of degree $p$ and $q$ w.r.t the knot sequences $U$ and $V$ in $[0,1]$ respectively. Since we are working in complex settings, we rewrite the parameterization by a complex function
\begin{equation}\label{B-spline complexrep}
    \begin{split}
f(z):=\sum_{i=0}^{m}\sum_{j=0}^{n}c_{ij}M_{i}^{p}(x)N_{j}^{q}(y),
    \end{split}
\end{equation}
where $c_{ij}=x_{ij}+\sqrt{-1}y_{ij}$, $i=0,1,\ldots,m$ and $j=0,1,\ldots,n$.
Since the boundary curves of the domain $\Omega$
are given, $c_{ij}$ is known for $i=0,1,\ldots,m$, $j=0,n$ and $j=0,1,\ldots,n$, $i=0,m$.

From the equation~(\ref{QCDefinition}), the Beltrami coefficient of $f$ can be computed as
\begin{equation}\label{ComputingBeltramiCoefficient}
    \begin{split}
\mu(f) = \frac{(a-b)+\sqrt{-1}(c+d)}{(a+b)+\sqrt{-1}(c-d)}
    \end{split}
\end{equation}
where
\begin{equation*}
\begin{split}
a = \sum_{i=0}^{m}\sum_{j=0}^{n}x_{ij}\frac{\partial M_{i}^{p}(x)}{\partial x}N_{j}^{q}(y),\quad b=\sum_{i=0}^{m}\sum_{j=0}^{n}y_{ij}M_{i}^{p}(x)\frac{\partial N_{j}^{q}(y)}{\partial y},\\
c=\sum_{i=0}^{m}\sum_{j=0}^{n}y_{ij}\frac{\partial M_{i}^{p}(x)}{\partial x}N_{j}^{q}(y),\quad d=\sum_{i=0}^{m}\sum_{j=0}^{n}x_{ij}M_{i}^{p}(x)\frac{\partial N_{j}^{q}(y)}{\partial y}
\end{split}
\end{equation*}

\subsection{Low-rank parameterization model}
As explained in Section~\ref{sec:QCMap}, the distortion of a quasi-conformal map $f$ is determined by its Beltrami coefficient $\mu(f)$, thus we formulate the parameterization problem as the following model
\begin{equation}\label{original_model}
    \begin{split}
\arg\min_{f} &\quad \int_{\hat{\Omega}}|\mu(f)|^2 \mathrm{d}z + \omega_1\int_{\hat{\Omega}}|\nabla \mu(f)|^2\mathrm{d}z +\omega_2 \rank(C)\\
      \mathrm{s.t.} &\quad \|\mu(f)\|_{\infty} < 1 \\
     &\quad c_{i0},c_{in},c_{0j},c_{mj}\ (i=0,1,\cdots,m,j=0,1,\cdots,n)\ \text{are given}
    \end{split}
\end{equation}
where $C=(c_{ij})_{(m+1)\times (n+1)}$ is a complex matrix whose elements are the coefficients of $f$ defined in (\ref{B-spline complexrep}), $\omega_1$ and $\omega_2$ are non-negative weights. The first term of the objective function aims to minimize the conformality distortion of $f$, the second term measures the smoothness of $f$ and the third term is the low-rank regularization term which tries to reduce the rank of $f$. In terms of the constraints, the first one guarantees that $f$ is locally bijective and the second one is the boundary conditions.

\subsection{Numerical algorithm}
Solving the optimization problem~(\ref{original_model}) for $f$ is challenging since it is highly nonlinear and nonconvex. Instead, we set $\nu:=\mu(f)$ as the auxiliary variable and replace the function $\rank(\cdot)$ with the nuclear norm $\|\cdot\|_{\ast}$ introduced in Section~\ref{sec:LRTA}. Thus we obtain the following optimization problem
\begin{equation}\label{alternative_model}
    \begin{split}
    \arg\min_{f,\nu} &\quad \int_{\hat{\Omega}}|\nu|^2 \mathrm{d}z + \omega_1\int_{\hat{\Omega}}|\nabla \nu|^2 \mathrm{d}z+\omega_2\|C\|_{\ast}\\
     \mathrm{s.t.} &\quad \nu=\mu(f)\\
                            &\quad \|\nu\|_{\infty}< 1\\
                   &\quad c_{i0},c_{in},c_{0j},c_{mj}\ (i=0,1,\cdots,m,j=0,1,\cdots,n)\ \text{are given}
    \end{split}
\end{equation}
The above problem is relaxed as

\begin{equation}\label{penalty_model}
    \begin{split}
    \arg\min_{f,\nu} &\quad \int_{\hat{\Omega}}|\nu|^2 \mathrm{d}z + \omega_1\int_{\hat{\Omega}}|\nabla \nu|^2 \mathrm{d}z+\omega_2\|C\|_{\ast}+\omega_3\int_{\hat{\Omega}}|\nu-\mu(f)|^2\mathrm{d}z\\
         \mathrm{s.t.} &\quad \|\nu\|_{\infty}< 1 \\
         &\quad c_{i0},c_{in},c_{0j},c_{mj}\ (i=0,1,\cdots,m,j=0,1,\cdots,n)\ \text{are given}
         \end{split}
\end{equation}
%The conventional penalty method increases the penalty parameter $\omega_3$ in each iteration to $\infty$. It iteratively solves a series of optimization problems with increasing $\omega_3$, whose solutions converge to the original constraint problem~(\ref{alternative_model})~\citep{nocedal2006numerical}. To enhance the efficiency of the algorithm, we fix $\omega_3$ to be large enough, in other words, we relax the constraint $\nu=\mu(f)$.
For large enough weight $\omega_3$, the optimal solution of the model~(\ref{penalty_model}) approximates the solution of~(\ref{alternative_model}), where $\nu$ is close enough to $\mu(f)$. To efficiently solve~(\ref{penalty_model}), we solve two sub-problems alternatively. More specifically, we set $\nu_0=0$ initially. Suppose $\nu_n$ is obtained at the $n$th iteration. Fixing $\nu=\nu_n$, we first minimize~(\ref{penalty_model}) for $f$ to obtain $f_n$. Then by fixing $f=f_n$, we obtain $\nu_{n+1}$ by minimizing~(\ref{penalty_model}) for $\nu$. The procedure runs until $\|\nu_{n+1}-\nu_{n}\|_{\infty}< \epsilon$ for a user-specified $\epsilon$. In the following, we will discuss the two sub-problems in detail.\\

\noindent\textbf{Problem 1}\label{problem1} Given $\nu$, find $f$ such that the following objective function is minimized
\begin{equation}\label{submodel1_original}
    \begin{split}
     \arg\min_{f} \quad &\omega_2\|C\|_{\ast}+\omega_3\int_{\hat{\Omega}}|\nu-\mu(f)|^2\mathrm{d}z\\
     \mathrm{s.t.}\quad &c_{i0},c_{in},c_{0j},c_{mj}\ (i=0,1,\cdots,m,j=0,1,\cdots,n)\ \text{are given}
    \end{split}
\end{equation}
 Problem~(\ref{submodel1_original}) is similar to the complex matrix completion problem~\citep{cai2010singular}.
 However, since $\mu(f)=f_{\bar z}/f_z$ is a rational B-spline function, the problem is still hard to solve. Instead we solve the following relaxed model
\begin{equation}\label{submodel1_new}
    \begin{split}
     \arg\min_{f} \quad \omega_2\|C\|_{\ast}+\omega_3\int_{\hat{\Omega}}|f_{\bar{z}}-\nu f_{z}|^2\mathrm{d}z + \lambda\|Pr(C)-y\|^2
    \end{split}
\end{equation}
where $\lambda$ is a large positive weight, $y\in \mathbb{C}^{2(m+n)}$ is the vector whose elements are comprised of
$c_{i0},c_{in}$,\linebreak $c_{0j},c_{mj}$ ($i=0,1,\cdots,m,j=0,1,\cdots,n$), and $Pr:\mathbb{C}^{(m+1)\times(n+1)}\to \mathbb{C}^{2(m+n)}$ is a linear operator that shapes the boundary elements of $C$ into a vector, i.e., $Pr(C)=(c_{00},,\cdots,c_{m0},c_{m1},\cdots,c_{mn},c_{m-1,n},\cdots,c_{0n}$,\linebreak $c_{0,n-1},\cdots,c_{01})^\text{T}$. Now~(\ref{submodel1_new}) is a convex optimization problem which can be solved by the alternating direction method of multipliers (ADMM) efficiently. The ADMM can be viewed as an attempt to blend the benefits of dual decomposition and augmented Lagrangian methods, and is used to solve constrained optimization problems with separable objective functions. The basic approach is outlined as follows. \\

\noindent\textbf{Variable splitting} Since the objective function in~(\ref{submodel1_new}) is the sum of three functions and one of which is dependent on the others, using variable splitting technique leads to the following constrained optimization problem
\begin{equation}\label{submodel1_admm}
    \begin{split}
     \arg\min_{f} \quad & \omega_2\|Z\|_{\ast}+\omega_3\int_{\hat{\Omega}}|f_{\bar{z}}-\nu f_{z}|^2\mathrm{d}z + \lambda\|Pr(C)-y\|^2\\
     \mathrm{s.t.} \quad &c=z
    \end{split}
\end{equation}
where $c$ is the vectorization of $C$, i.e., $c=(c_{00},\cdots,c_{m0},c_{01},\cdots,c_{m1},\cdots,c_{0n},\cdots,c_{mn})^\text{T}$, $Z$ is an auxiliary matrix of the same size as $C$, and $z$ is the vectorization of $Z$. \\

\noindent\textbf{Augmented Lagrangian}
One typical way for solving~(\ref{submodel1_admm}) is to use the augmented Lagrangian scheme. In our problem, the augmented Lagrangian function is defined as
\begin{equation}\label{AL}
    \begin{split}
     \mathcal{L}_{\rho}(c,Z,\eta)&= \omega_2\|Z\|_{\ast}+\omega_3\int_{\hat{\Omega}}|f_{\bar{z}}-\nu f_{z}|^2\mathrm{d}z + \lambda\|Pr(C)-y\|^2\\
     &\quad+\mathbf{Re}(\eta^\mathrm{H}(c-z))+\frac{\rho}{2}\|c-z\|^2
    \end{split}
\end{equation}
where $\mathbf{Re}(z)$ is the real part of the complex number z, $\eta$ is a vector of Lagrangian multiplier corresponding to the constraint $c=z$, and $\rho>0$ is the penalty parameter.
Now the ADMM algorithm can be outlined as follows
\begin{algorithm}[H]
\begin{algorithmic}[1]
	\caption{The ADMM algorithm} \label{alg:ADMM}
    \REQUIRE ~~\\ %Input
    $\omega_2,\omega_3,\lambda,\rho>0$, and
    $\text{initial\ values for } Z^0, \eta^0$.\\
    \ENSURE ~~\\ %Output
    an optimal $c^{\ast}$\\

    \STATE {$t \leftarrow 0$}
	\REPEAT
		\STATE {$c^{t+1}=\arg\min\limits_{c}\mathcal{L}_{\rho}(c,Z^t,\eta^t)$}
		\STATE {$Z^{t+1}=\arg\min\limits_{Z}\mathcal{L}_{\rho}(c^{t+1},Z,\eta^t)$}
		\STATE {$\eta^{t+1}=\eta^{t}+\rho(c^{t+1}-z^{t+1})$}
        \STATE {$t \leftarrow t+1$}
	\UNTIL {stopping criterion is satisfied.}
\end{algorithmic}
\end{algorithm}
\noindent {\bf $c$-subproblem}\; The subproblem for $c$ is
\begin{equation}\label{csubproblem}
\begin{aligned}
    &\arg\min\limits_{c}\mathcal{L}_{\rho}(c,Z^t,\eta^t)=\omega_3\int_{\hat{\Omega}}|f_{\bar{z}}-\nu f_{z}|^2\mathrm{d}z + \lambda\|Pr(C)-y\|^2
     +\mathbf{Re}((\eta^t)^\mathrm{H}(c-z^t))+\frac{\rho}{2}\|c-z^t\|^2
\end{aligned}
\end{equation}
This is a quadratic optimization problem and the solution can be obtained by solving a sparse and symmetric linear system of equations. The preconditioned conjugate gradient method with incomplete Cholesky factorization is applied in our algorithm.\\

\noindent{\bf $Z$-subproblem\;} The subproblem for $Z$ is
\begin{equation}\label{zsubproblem}
\begin{aligned}
    &\arg\min\limits_{Z}\mathcal{L}_{\rho}(c^{t+1},Z,\eta^t)=\omega_2\|Z\|_{\ast}+\mathbf{Re}((\eta^t)^\mathrm{H}(c^{t+1}-z))+\frac{\rho}{2}\|c^{t+1}-z\|^2
\end{aligned}
\end{equation}
which has the following closed form solution~\citep{cai2010singular}:
\begin{equation}\label{wsubproblem}
    Z^{t+1}=\mathrm{prox}^{\mathrm{tr}}_{{\omega_2}/\rho}(c^{t+1}+\eta^t/{\rho}), \;
\end{equation}
Note that the argument $c^{t+1}+\eta^t/{\rho}$ must be converted into a matrix of the same size as $Z^t$. Here the proximal operator $\mathrm{prox}^{\mathrm{tr}}_{{\omega_2}/\rho}$ can be considered as a shrinkage operation on the singular values and is defined as follows
\begin{equation}\label{shrinkage}
\mathrm{prox}^{\mathrm{tr}}_{{\omega_2}/\rho}(Y)=U\mathrm{max}(S-{\omega_2}I/\rho,0)V^\mathrm{H},
\end{equation}
where $Y = USV^\mathrm{H}$ is the singular value decomposition (SVD) of $Y$, and the max operation is taken element-wise. Please refer to \citep{cai2010singular} for the detailed derivation.

In our implementation, $Z^0$ and $\eta^0$ are set as zero, and the stopping criterion is that the value of $c$ has small change or the maximum number of  iterations reaches.\\

\noindent\textbf{Problem 2}
Given a mapping $f$ from $\hat{\Omega}$ to $\Omega$, $\mu(f)$ can be computed by~(\ref{ComputingBeltramiCoefficient}), and the problem~(\ref{penalty_model}) reduces to the following model
\begin{equation}\label{submodel2_original}
    \begin{split}
     \arg\min_{\nu}\quad &\int_{\hat{\Omega}}|\nu|^2 \mathrm{d}z + \omega_1\int_{\hat{\Omega}}|\nabla \nu|^2 \mathrm{d}z+\omega_3\int_{\hat{\Omega}}|\nu-\mu(f)|^2\mathrm{d}z\\
     \mathrm{s.t.}\quad &\|\nu\|_{\infty}< 1
    \end{split}
\end{equation}
Let
\begin{equation}\label{nuexpression}
\nu=\sum_{i=0}^{\tilde{m}}\sum_{j=0}^{\tilde{n}}{\tilde c}_{ij}M_{i}^p(x)N_{j}^q(y),
\end{equation}
where ${\tilde c}_{ij}=\tilde{x}_{ij}+\sqrt{-1}\tilde{y}_{ij}$, and $M_{i}^p(x)$ and $N_{j}^q(y)$ are the B-spline basis functions defined in~(\ref{B-spline rep}). For the simplicity of computation, the constraint $\|\nu\|_{\infty}< 1$ in the above optimization problem is replaced by
\begin{equation}\label{reformulatedconstraint}
    \begin{split}
    -\frac{\sqrt{2}}{2}<\tilde{x}_{ij}<\frac{\sqrt{2}}{2},\ -\frac{\sqrt{2}}{2}<\tilde{y}_{ij}<\frac{\sqrt{2}}{2}, \quad i=0,1,\cdots,\tilde{m},j=0,1,\cdots,\tilde{n}
    \end{split}
\end{equation}
Then~(\ref{submodel2_original}) becomes a quadratic optimization problem which can be easily solved.

\subsection{Post-processing}
\label{sec:postprocessing}
\begin{figure}[!htbp]
    \centering
        \includegraphics[width=0.65\textwidth]{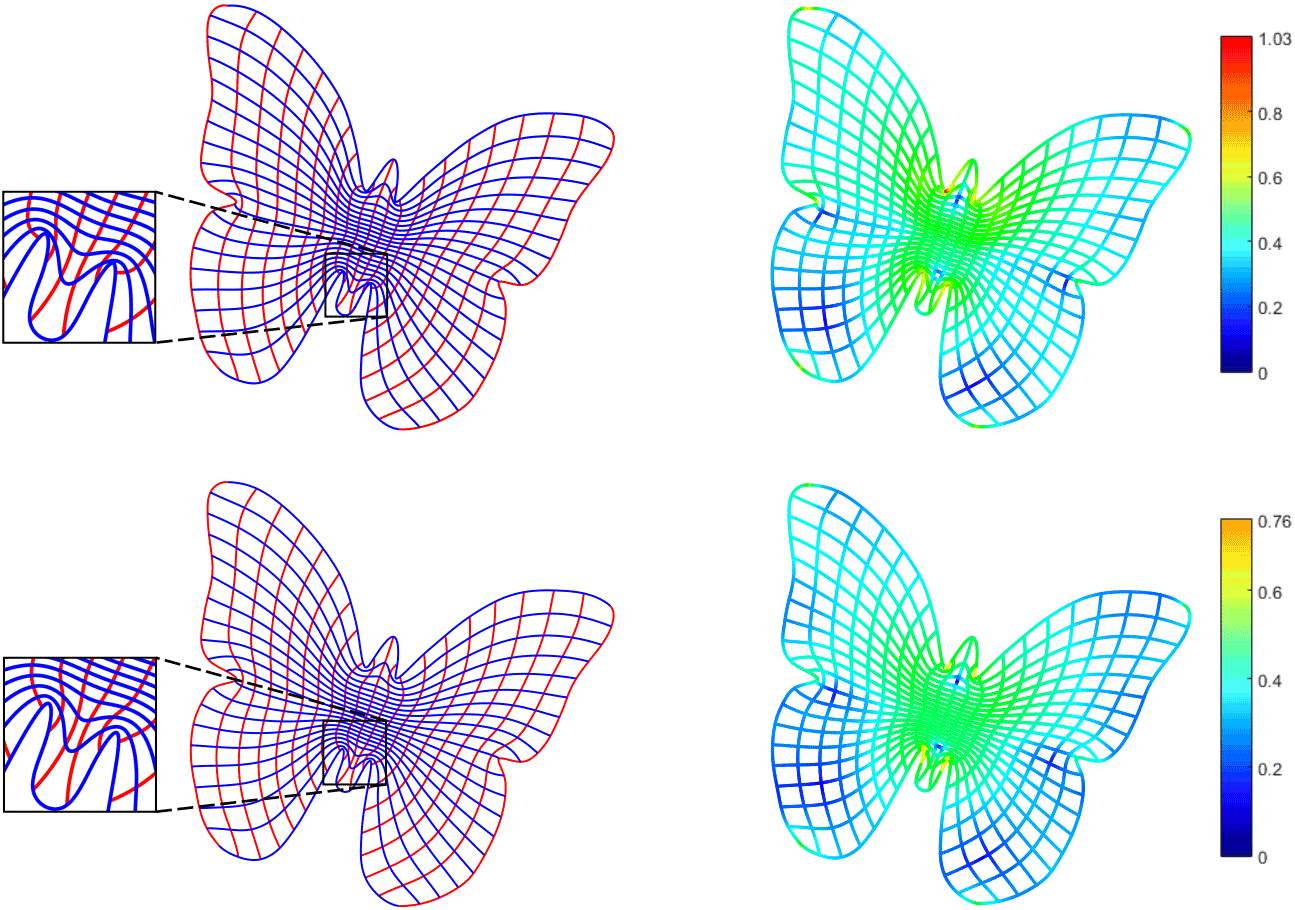}
     \caption{Comparison of parameterization results with (bottom row) and without (top row) post-processing step. The left column shows the iso-parametric curves with some parts enlarged and the right one shows the colormaps of $|\mu(f)|$ (i.e., angular distortion). }
    \label{postprocessingcomparison}
\end{figure}

The above algorithm iteratively solves two sub-problems to obtain two sequences of complex functions $\{\nu_k\}$ and $\{f_k\}$. In order to accelerate the convergence of the algorithm, we add a weight into the second term of problem~(\ref{submodel1_original}) after $t_0$ iterations, where $t_0$ satisfies $\|\mu(f_{t_0})\|_{\infty}-1<\epsilon_0$ for a threshold $\epsilon_0$, which leads to the following problem
\begin{equation}\label{postprocessing}
    \begin{split}
     \arg\min_{f}\quad &\omega_2\|C\|_{\ast}+\omega_3\int_{\hat{\Omega}}\omega|\nu-\mu(f)|^2\mathrm{d}z\\
     \mathrm{s.t.} \quad &c_{i0},c_{in},c_{0j},c_{mj}\ (i=0,1,\cdots,m,j=0,1,\cdots,n)\ \text{are given}
    \end{split}
\end{equation}
where the weight $\omega=1/((1-|\mu(f_{t_0})|)^2+\delta)$ and $\delta$ is a threshold which helps to avoid division by zero. The problem (\ref{postprocessing}) can be solved in the same way as the problem (\ref{submodel1_original}). From the numerical examples, we can see that this post-processing step is essential and effective, see Fig.~\ref{postprocessingcomparison} for a comparison result.

\smallskip

Now the overall algorithm of our parameterization method  is summarized in Algorithm~\ref{alg:LRPA}.
\begin{algorithm}[H]
\begin{algorithmic}[1]
	\caption{The low-rank parameterization algorithm} \label{alg:LRPA}
    \REQUIRE ~~\\ %Input
    B-spline representations of four boundary curves of the domain $\Omega$, \\and the parameters $\omega_1,\omega_2,\omega_3,\lambda,\epsilon,\epsilon_0,\delta$.
    \ENSURE ~~\\ %Output
   A low-rank quasi-conformal map $f$ from the unit square $\hat{\Omega}$ to $\Omega$.\\
    \STATE {$\nu_0=0,f_0=10,k=0$}
	\REPEAT
      \IF{$\|\mu(f_k)\|_{\infty}-1\geq\epsilon_0$}
      \STATE {Fix $\nu_k$, and solve the optimization problem (\ref{submodel1_original}) to obtain $f_k$  using Algorithm~\ref{alg:ADMM},\\
      $\omega=1/((1-|\mu(f_k)|)^2+\delta)$;}
        \ELSE
         \STATE {Fix $\nu_k$ and $w$, and solve (\ref{postprocessing}) to obtain $f_k$  using Algorithm~\ref{alg:ADMM};}
       \ENDIF
		\STATE {Fix $f_k$,  and solve (\ref{submodel2_original}) to obtain $\nu_{k+1}$;}
        \STATE {$k \leftarrow k+1$;}
	\UNTIL {$\|\nu_{k+1}-\nu_{k}\|_{\infty}<\epsilon$.}
\end{algorithmic}
\end{algorithm}

\begin{remark}
Our mathematical model~(\ref{original_model}) and the registration model presented in~\citep{lam2014landmark} both obtain a diffeomorphism via quasi-conformal mapping. However, there are several differences between the two methods. Firstly, we not only want to find a quasi-conformal map with low distortion but also add a low-rank regularization term in~(\ref{original_model}) to make the rank of the map as low as possible. Secondly, in~\citep{lam2014landmark}, the map $f$ is represented in a discrete form, while it is expressed in a continuous form, i.e., tensor product B-splines in our work. Finally, the second constraint in the model~(\ref{original_model}) is ignored in~\citep{lam2014landmark}, which can not guarantee the bijectivity of $f$.
\end{remark}

\section{Results and discussions}
\label{sec:result}
In this section, we demonstrate some examples to show the effectiveness of our parameterization method by comparing it with several state-of-the-art methods. The application of our parameterization in solving numerical PDEs is also provided.
\subsection{Implementation details}
We implement our algorithm on a PC with a quad-core Intel i5 @3.1GHz and 8GB of RAM using C++ and MATLAB. There are several parameters for setting. Most of them are set as default values, e.g., the penalty parameter $\rho$ is typically set to be $1$, the threshold $\epsilon_0$ and $\delta$ in Section~\ref{sec:postprocessing} are set as $0.05$ and $0.0001$ respectively, and the weight $\lambda$ in~(\ref{submodel1_new}) is set to be $1000$. We use bicubic uniform B-splines to represent the map $f$ and the auxiliary variable $\nu$ (i.e., $p=q=3$ in~(\ref{B-spline complexrep}) and~(\ref{nuexpression})). Unless specified, the knot parameters in~(\ref{B-spline complexrep}) and~(\ref{nuexpression}) are chosen as $m=n=24$ and $\tilde{m}=\tilde{n}=46$ respectively in our examples, which is proven to work well.

\begin{figure}[!htbp]
    \centering
    \subfigure[]{
        \label{DifferentModels:w1}
        \includegraphics[width=0.24\textwidth]{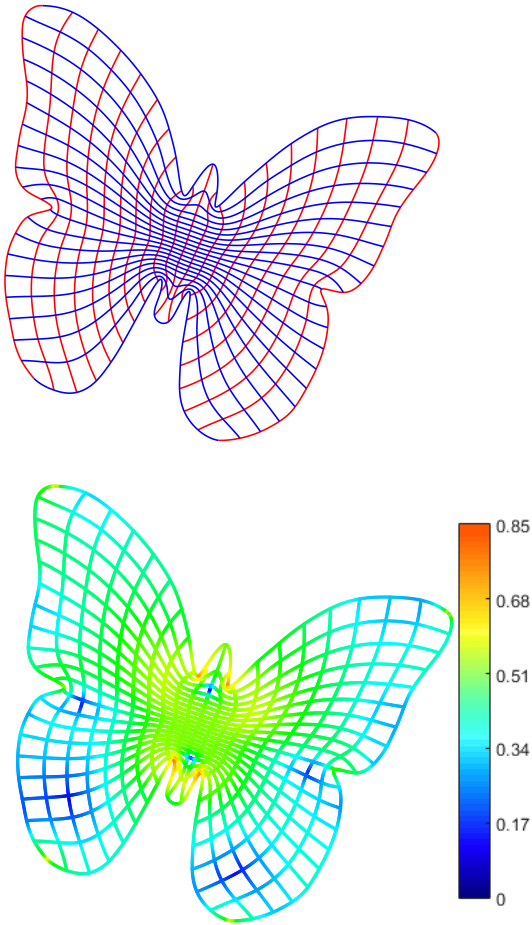}
    }
    \hspace{-0.025\textwidth}
    \subfigure[]{
        \label{DifferentModels:w2}
        \includegraphics[width=0.24\textwidth]{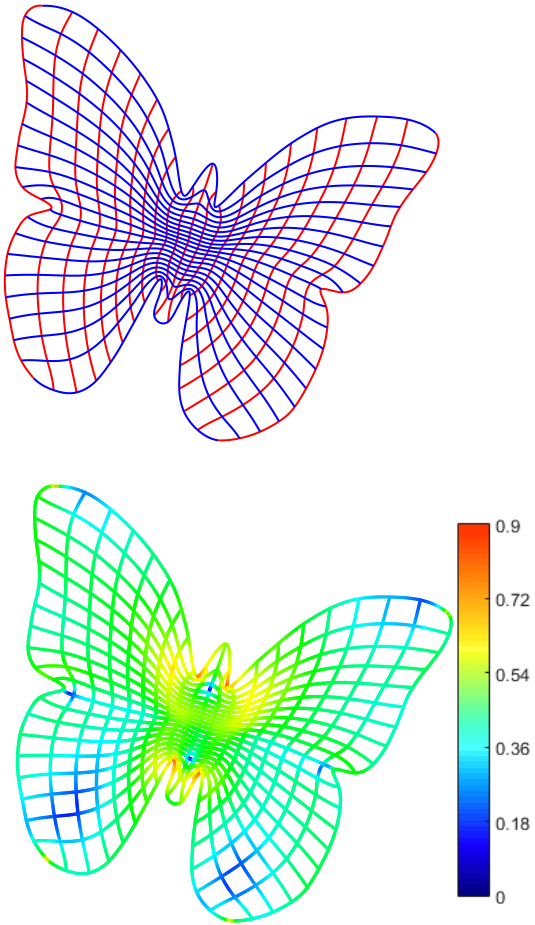}
    }
    \hspace{-0.025\textwidth}
     \subfigure[]{
        \label{DifferentModels:w3}
        \includegraphics[width=0.24\textwidth]{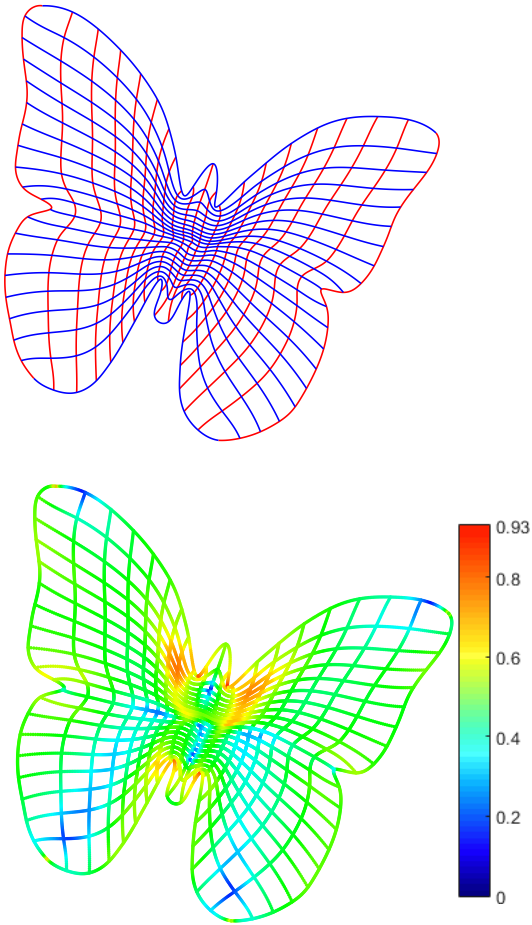}
    }
    \hspace{-0.025\textwidth}
    \subfigure[]{
        \label{DifferentModels:w4}
        \includegraphics[width=0.24\textwidth]{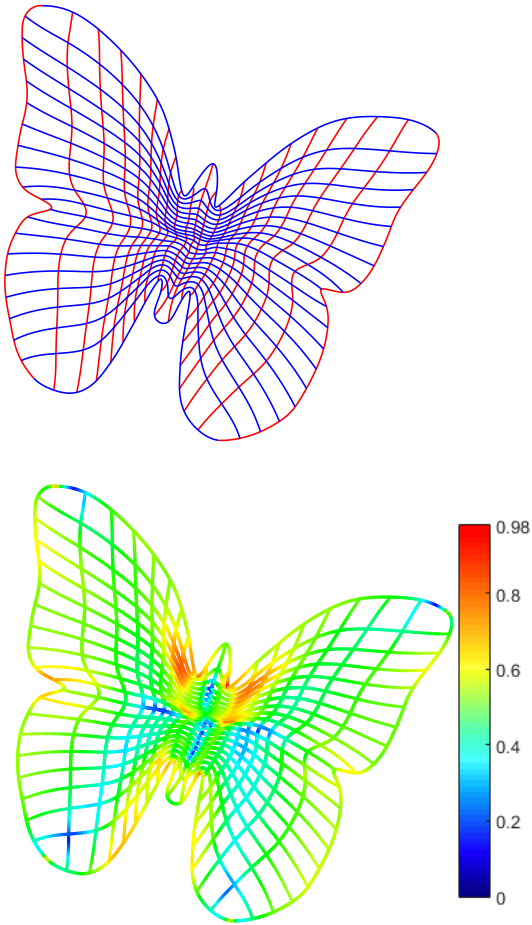}
    }
	\caption{Parameterization of the Butterfly model using various weights $\omega_2$ while leaving other parameters fixed. The top and bottom rows show the iso-parametric curves and the colormaps of $|\mu(f)|$ respectively. \subref{DifferentModels:w1} $\omega_2=1.5$, $\rank(f)=11$;
    \subref{DifferentModels:w2} $\omega_2=5.5$, $\rank(f)=7$;\subref{DifferentModels:w3} $\omega_2=7.5$, $\rank(f)=5$;
     \subref{DifferentModels:w4} $\omega_2=10.0$, $\rank(f)=4$.}
    \label{DifferentModels}
\end{figure}

There are three weights $\omega_1$, $\omega_2$ and $\omega_3$ in the mathematical model~(\ref{penalty_model}). The weight $\omega_1$ controls the smoothness of $f$ and we typically set $\omega_1\in[3,5.5]$. The penalty weight $\omega_3$ is used to control the difference between $\nu$ and $\mu(f)$ and is set to be 100 in practice. The weight $\omega_2$ can be used to balance the the rank of the map $f$ and parameterization quality. Clearly, larger $\omega_2$ can reduce the rank of $f$ while smaller $\omega_2$ leads to parameterization results of higher quality. We observe that $\omega_2\in[4.5,6]$ provides a good compromise between the rank and the quality. Fig.~\ref{DifferentModels} provides an illustrating example.

\subsection{Parameteriation results}

In the following, we present some examples to demonstrate the low rank and low distortion properties of our parameterization method.

\subsubsection{Low rank}
\begin{figure}[!htbp]
    \centering
    \subfigure[Bear]{
        \label{ParametrizationResults:Bear}
        \includegraphics[width=0.25\textwidth]{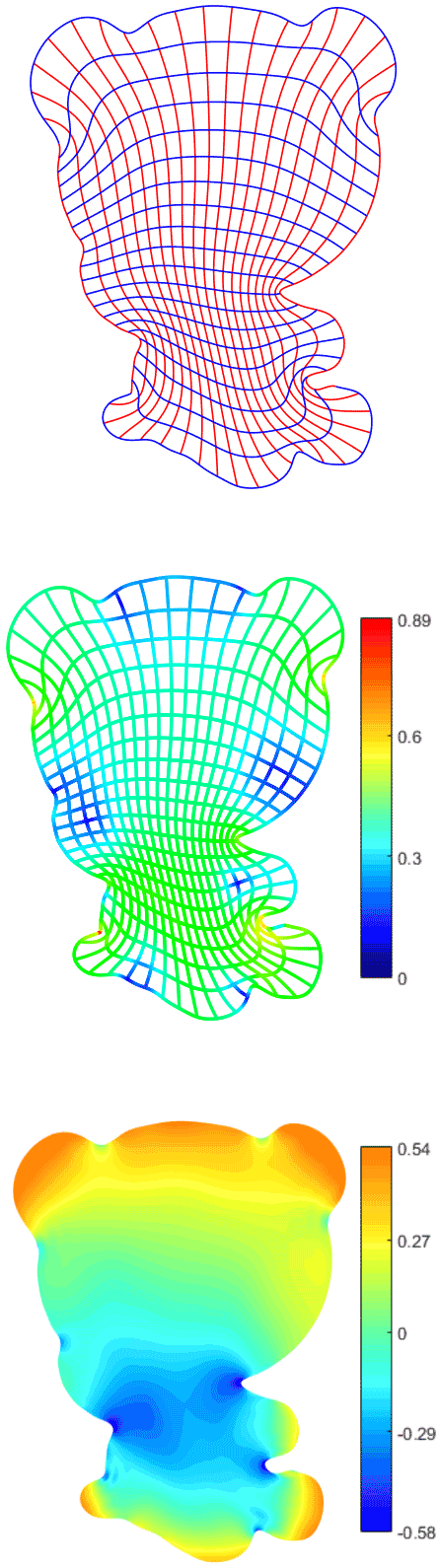}
    }
    \hspace{-0.001\textwidth}
    \subfigure[Sheep]{
        \label{ParametrizationResults:Sheep}
        \includegraphics[width=0.25\textwidth]{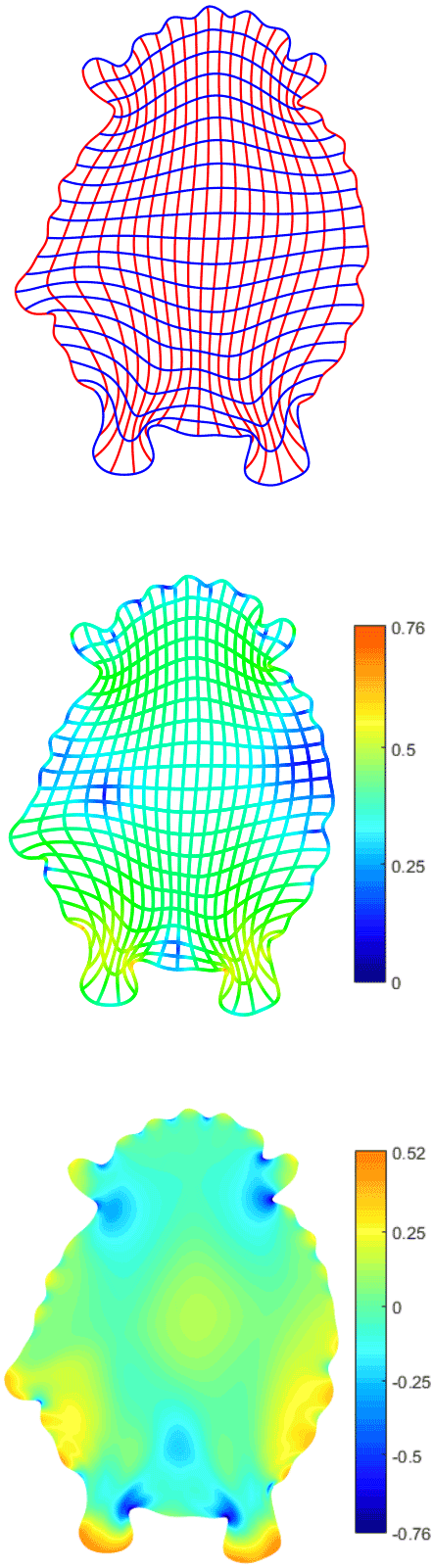}
    }
    \hspace{-0.015\textwidth}
    \subfigure[Jigsaw]{
        \label{ParametrizationResults:Jigsaw}
        \includegraphics[width=0.4\textwidth]{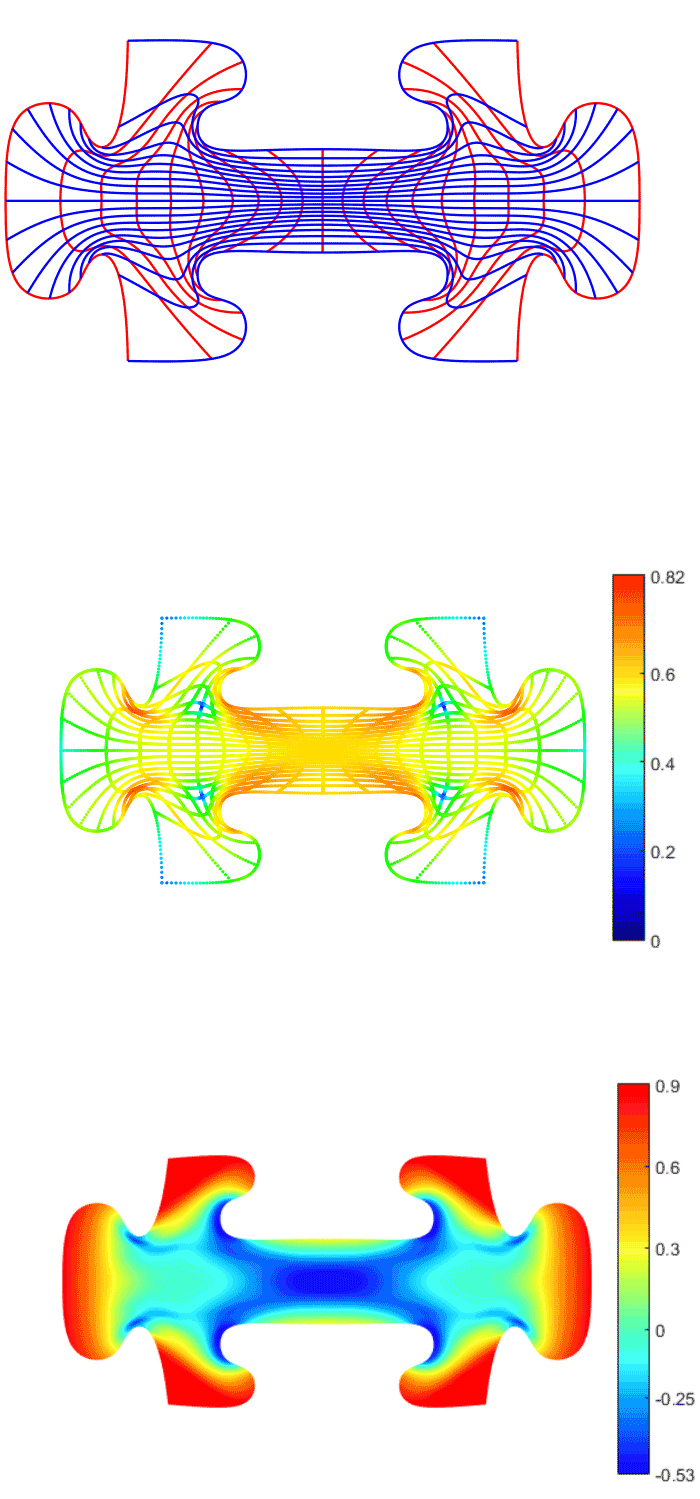}
    }
	\caption{Parameterization results of different models by our method. \subref{ParametrizationResults:Bear} the Bear model, \subref{ParametrizationResults:Sheep} the Sheep model, \subref{ParametrizationResults:Jigsaw} the Jigsaw model. The top row shows the iso-parametric curves, the middle and bottom rows show the colormaps of $|\mu(f)|$ and $\log_{10}|J_s(f)|$ respectively. Note that the optimal values of $|\mu(f)|$ and $\log_{10}|J_s(f)|$ are both $0$.}
\label{ParametrizationResults}
\end{figure}

To demonstrate the superiority of our method in terms of the rank of the map, we provide a comparison with several state-of-the-art parameterization methods: nonlinear optimization method~\citep{xu2011parameterization}, variational harmonic method~\citep{xu2013constructing}, the Teichm{\"u}ller mapping method (T-Map)~\citep{nian2016planar} and the low-rank spline interpolation method (low-rank spline)~\citep{juttler2017low}. To have a fair comparison, the number of knots in the B-spline representation (\ref{B-spline rep}) are chosen to be the same ($m=n=24$) for these methods. The rank of the map $f$ is numerically computed as the number of singular values of the complex matrix $C$ which are greater than a user-specified threshold ($10^{-5}$ in our experiment). \autoref{rankcomparision} shows the statistics of our experiments. Besides the low-rank spline method which sacrifices the parameterization quality, our method significantly outperforms other state-of-the-art parameterization methods in terms of the rank. Some of the parameterization results are shown in Fig~\ref{ParametrizationResults}. As described in Section~\ref{sec:rankRfunction}, owing to the low-rank property of our method, the map $f$ can be represented in a sum  of the product of univariate spline functions with a small number of terms, which helps to speed up the assembly process in IGA without sacrificing  the overall accuracy of the simulation, see Section~\ref{solvingpdes} for some examples.

\begin{table*}[!htbp]
	\begin{center}
		\scriptsize
        \renewcommand{\arraystretch}{1.2}
        \def\temptablewidth{0.74\textwidth}
        {\rule{\temptablewidth}{1.0pt}}
		\begin{tabular}{|c|c|c|c|c|c|c|}\hline
			%\multicolumn{1}{c}{\multirow{3}{*}{Model}}
			\multirow{2}{*}{Model}&\multirow{2}{*}{Quantity} &Variational&Low-rank&Nonlinear &\multirow{2}{*}{T-Map}&\multirow{2}{*}{Ours}\\
            &&harmonic&spline&optimization&&\\
			\hline \hline
            %\multirow{2}{*}{Bunny(Fig~\ref{BunnyDifferentWeight})}
			\multirow{3}{*}{Sheep (Fig.~\ref{ParametrizationResults})}&$\|\mu(f)\|_{\infty}$&45.40&26.92 &0.99&1.03&0.76\\
            \cline{2-7}
            &$\min(J_s(f))$&-1.52&-2.59&0.06&0.11&0.17\\
            \cline{2-7}
            &$\max(J_s(f))$&4.38&4.67&10.28&4.24&3.33\\
			\cline{2-7}
			&$\rank(f)$&24&5&22&25&7\\
            \hline
            \multirow{3}{*}{Bear (Fig.~\ref{ParametrizationResults})}&$\|\mu(f)\|_{\infty}$&75.91&80.36&0.98&0.96&0.89\\
			\cline{2-7}
			&$\min(J_s(f))$&-1.25&-1.89&0.11&0.21&0.26\\
            \cline{2-7}
            &$\max(J_s(f))$&5.22&3.86&11.49&4.85&3.52\\
            \cline{2-7}
            &$\rank(f)$&25&5&24&25&6\\
            \hline
			\multirow{3}{*}{Dolphin (Fig.~\ref{DolphinInjectivityComparison})}&$\|\mu(f)\|_{\infty}$&2.51&5.86&0.99&0.98&0.92\\
			\cline{2-7}
            &$\min(J_s(f))$&-3.63&-4.32&0.05&0.10&0.23\\
            \cline{2-7}
            &$\max(J_s(f))$&5.26&4.93&11.57&6.07&5.88\\
            \cline{2-7}
            &$\rank(f)$&24&5&21&25&8\\
            \hline
			\multirow{3}{*}{Rabbit (Fig.~\ref{RabbitInjectivityComparison})}&$\|\mu(f)\|_{\infty}$&20.76&62.67&1.00&1.45&0.86\\
            \cline{2-7}
            &$\min(J_s(f))$&-2.95&-4.12&0.08&-0.94&0.19\\
            \cline{2-7}
            &$\max(J_s(f))$&4.37&5.15&8.34&6.18&4.22\\
            \cline{2-7}
            &$\rank(f)$&21&5&24&25&8\\
            \hline
            \multirow{3}{*}{Butterfly (Fig.~\ref{ButterflyInjectivityComparison})}&$\|\mu(f)\|_{\infty}$&14.80&4.12&0.96&0.8&0.76\\
			\cline{2-7}
            &$\min(J_s(f))$&-0.52&-0.66&0.04&0.02&0.15\\
            \cline{2-7}
            &$\max(J_s(f))$&3.67&3.01&7.58&3.82&3.51\\
            \cline{2-7}
            &$\rank(f)$&25&5&23&25&7\\
            \hline
            \multirow{3}{*}{Jigsaw (Fig.~\ref{ParametrizationResults})}&$\|\mu(f)\|_{\infty}$&76.75&484.25&0.91&0.8&0.82\\
			\cline{2-7}
            &$\min(J_s(f))$&-1.92&-8.85&0.04&0.22&0.29\\
            \cline{2-7}
            &$\max(J_s(f))$&21.58&31.24&25.75&8.84&8.02\\
            \cline{2-7}
            &$\rank(f)$&18&2&25&25&9\\
            \hline
		\end{tabular}
        {\rule{\temptablewidth}{1.0pt}}
		\caption{Comparisons of the distortions and ranks between our method and nonlinear optimization method, variational harmonic method, T-map method and the low-rank spline interpolation method. The distortion includes the angular distortion $|\mu(f)|$ and the area distortion $J_s(f)$. Here $\min(J_s(f))$, $\max(J_s(f))$ are the minimum and maximum area distortions respectively.}
		\label{rankcomparision}	
	\end{center}
\end{table*}

\subsubsection{Local injectivity}

\begin{figure}[!htbp]
    \centering
    \subfigure[Variational harmonic]{
        \label{RabbitInjectivityComparison:VH}
        \includegraphics[width=0.198\textwidth]{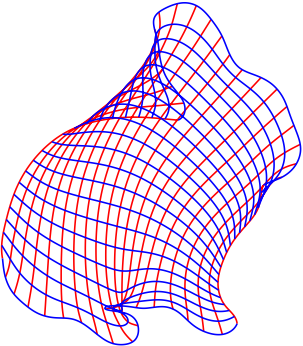}
    }
    \hspace{-0.029\textwidth}
    \subfigure[Low-rank spline]{
        \label{RabbitInjectivityComparison:LR}
        \includegraphics[width=0.198\textwidth]{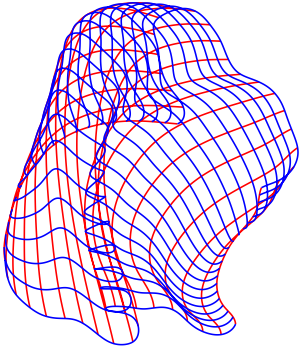}
    }
    \hspace{-0.029\textwidth}
    \subfigure[Nonlinear optimization]{
        \label{RabbitInjectivityComparison:NO}
        \includegraphics[width=0.198\textwidth]{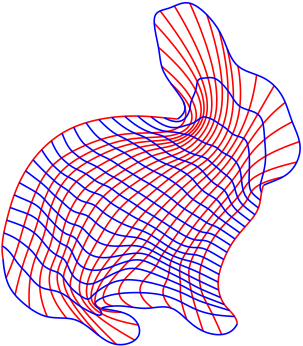}
    }
    \hspace{-0.029\textwidth}
    \subfigure[T-Map]{
        \label{RabbitInjectivityComparison:T-Map}
        \includegraphics[width=0.198\textwidth]{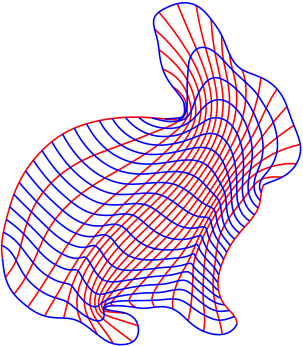}
    }
    \hspace{-0.029\textwidth}
    \subfigure[Ours]{
        \label{RabbitInjectivityComparison:Ours}
        \includegraphics[width=0.198\textwidth]{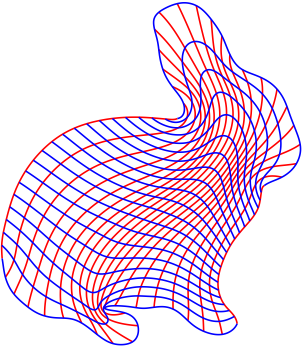}
    }
	\caption{Parameterization of the Rabbit model using different methods: \subref{RabbitInjectivityComparison:VH} variational harmonic, \subref{RabbitInjectivityComparison:LR} low-rank spline, \subref{RabbitInjectivityComparison:NO} nonlinear optimization, \subref{RabbitInjectivityComparison:T-Map} T-map method and \subref{RabbitInjectivityComparison:Ours} our method.}
    \label{RabbitInjectivityComparison}
\end{figure}

\begin{figure}[!htbp]
    \centering
    \subfigure[Variational harmonic]{
        \label{ButterflyInjectivityComparison:VH}
        \includegraphics[width=0.197\textwidth]{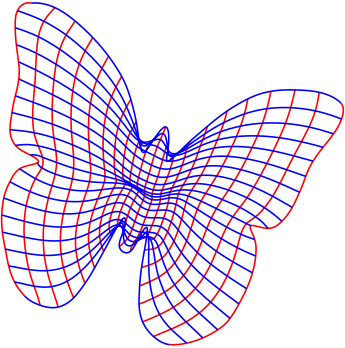}
    }
    \hspace{-0.028\textwidth}
    \subfigure[Low-rank spline]{
        \label{ButterflyInjectivityComparison:LR}
        \includegraphics[width=0.197\textwidth]{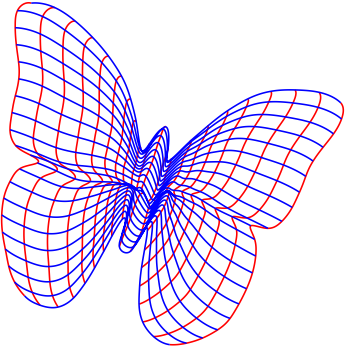}
    }
    \hspace{-0.028\textwidth}
    \subfigure[Nonlinear optimization]{
        \label{ButterflyInjectivityComparison:NO}
        \includegraphics[width=0.197\textwidth]{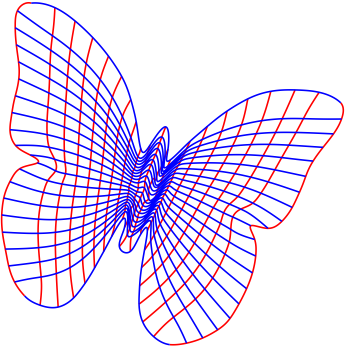}
    }
    \hspace{-0.028\textwidth}
    \subfigure[T-Map]{
        \label{ButterflyInjectivityComparison:T-Map}
        \includegraphics[width=0.197\textwidth]{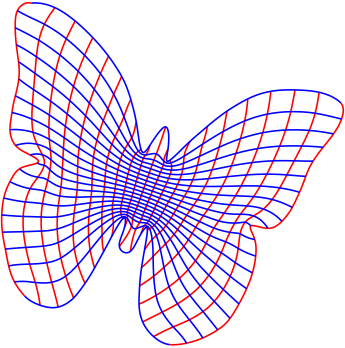}
    }
    \hspace{-0.028\textwidth}
    \subfigure[Ours]{
        \label{ButterflyInjectivityComparison:Ours}
        \includegraphics[width=0.197\textwidth]{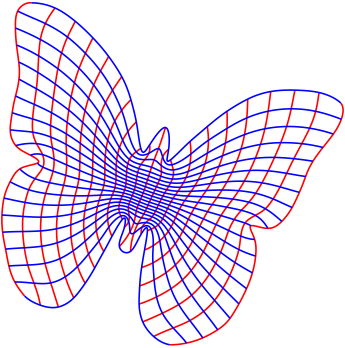}
    }
	\caption{Parameterization of the Butterfly model using different methods: \subref{RabbitInjectivityComparison:VH} variational harmonic, \subref{ButterflyInjectivityComparison:LR} low-rank spline, \subref{ButterflyInjectivityComparison:NO} nonlinear optimization, \subref{ButterflyInjectivityComparison:T-Map} T-map method and \subref{ButterflyInjectivityComparison:Ours} our method.}
\label{ButterflyInjectivityComparison}
\end{figure}

\begin{figure}[!htbp]
    \centering
    \subfigure[Variational harmonic]{
        \label{DolphinInjectivityComparison:VH}
        \includegraphics[width=0.197\textwidth]{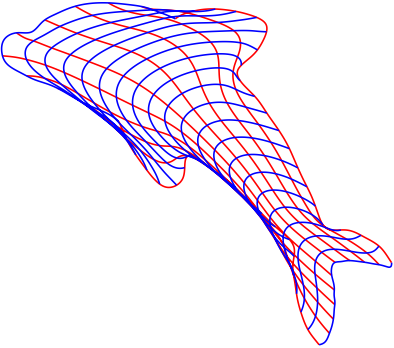}
    }
    \hspace{-0.029\textwidth}
    \subfigure[Low-rank spline]{
        \label{DolphinInjectivityComparison:LR}
        \includegraphics[width=0.197\textwidth]{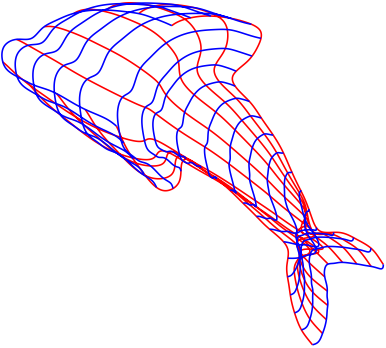}
    }
    \hspace{-0.029\textwidth}
    \subfigure[Nonlinear optimization]{
        \label{DolphinInjectivityComparison:NO}
        \includegraphics[width=0.197\textwidth]{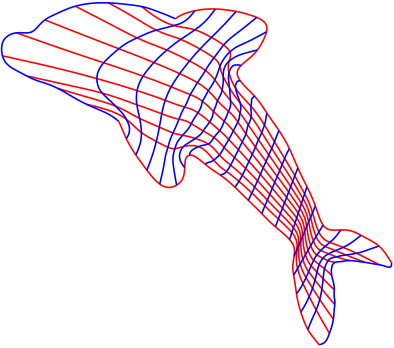}
    }
    \hspace{-0.029\textwidth}
    \subfigure[T-Map]{
        \label{DolphinInjectivityComparison:T-Map}
        \includegraphics[width=0.197\textwidth]{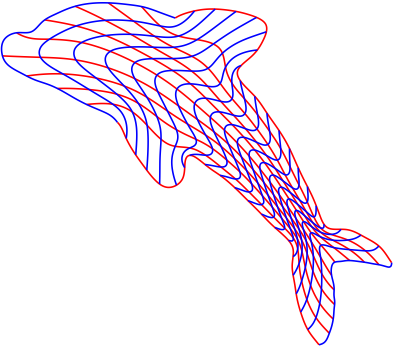}
    }
    \hspace{-0.029\textwidth}
    \subfigure[Ours]{
        \label{DolphinInjectivityComparison:Ours}
        \includegraphics[width=0.197\textwidth]{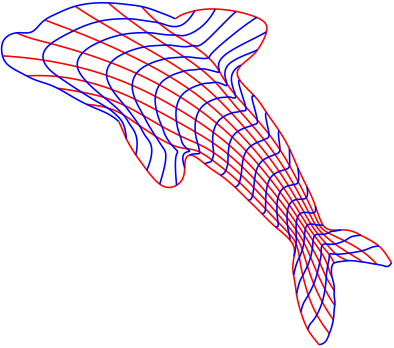}
    }
	\caption{Parameterization of the Dolphin model using different methods: \subref{DolphinInjectivityComparison:VH} variational harmonic, \subref{DolphinInjectivityComparison:LR} low-rank spline, \subref{DolphinInjectivityComparison:NO} nonlinear optimization, \subref{DolphinInjectivityComparison:T-Map} T-map method and \subref{DolphinInjectivityComparison:Ours} our method.}
    \label{DolphinInjectivityComparison}
\end{figure}

Fig.~\ref{RabbitInjectivityComparison}, Fig.~\ref{ButterflyInjectivityComparison} and Fig.~\ref{DolphinInjectivityComparison} depict the parameterization results of the rabbit, the butterfly and the dolphin by different methods respectively. We observe that the variational harmonic method and the low-rank spline method have many self-intersections in the concave regions, the nonlinear optimization method produces non-injective mapping in the butterfly model, the T-map method is not injective in some regions in the Rabbit model, e.g. in the ear of the rabbit, while our method is always injective in these examples.

\subsubsection{Distortion}

Besides injectivity, the map distortion (including angular distortion described by $|\mu(f)|$ and the area distortion represented by $J_s(f)$) is an important criteria to measure the quality of the parameterization.
In our experiments, to measure the area distortion of a map, we firstly uniformly subdivide the parametric domain $\hat{\Omega}$ into $M\times N$ sub-rectangles $\{\hat{\Omega}_{ij}\}$ ($i=1,\ldots,M$, $j=1,\ldots,N$),  then the area distortion over the sub-rectangle $\hat{\Omega}_{ij}$, denoted as $J_s(f)|_{\hat{\Omega}_{ij}}$, is calculated as follows
\begin{equation}\label{areadistortion}
    \begin{split}
    J_s(f)|_{\hat{\Omega}_{ij}} = \frac{\int_{\hat{\Omega}_{ij}}J_s(f)\mathrm{d}x\mathrm{d}y}{A_{\hat{\Omega}_{ij}}}
    \end{split}
\end{equation}
where $A_{\hat{\Omega}_{ij}}$ is the area of $\hat{\Omega}_{ij}$.
%The statistics in~\autoref{rankcomparision} indicate that our method provides lower angular and area distortions than other methods.

\begin{figure}[!htbp]
    \centering
    \subfigure[Nonlinear optimization]{
        \label{RabbitmufareaComparison:NO}
        \includegraphics[width=0.25\textwidth]{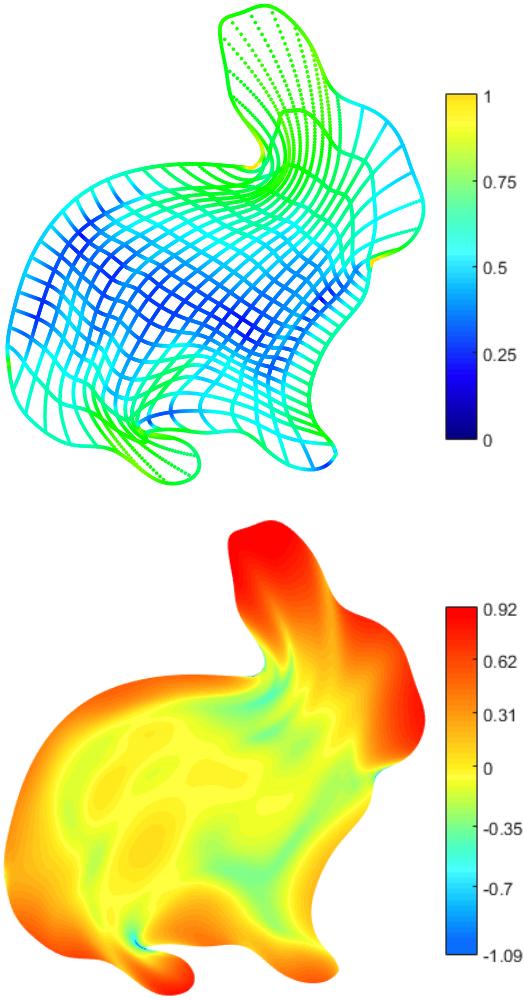}
    }
    \hspace{0.02\textwidth}
    \subfigure[T-map]{
        \label{RabbitmufareaComparison:T-Map}
        \includegraphics[width=0.25\textwidth]{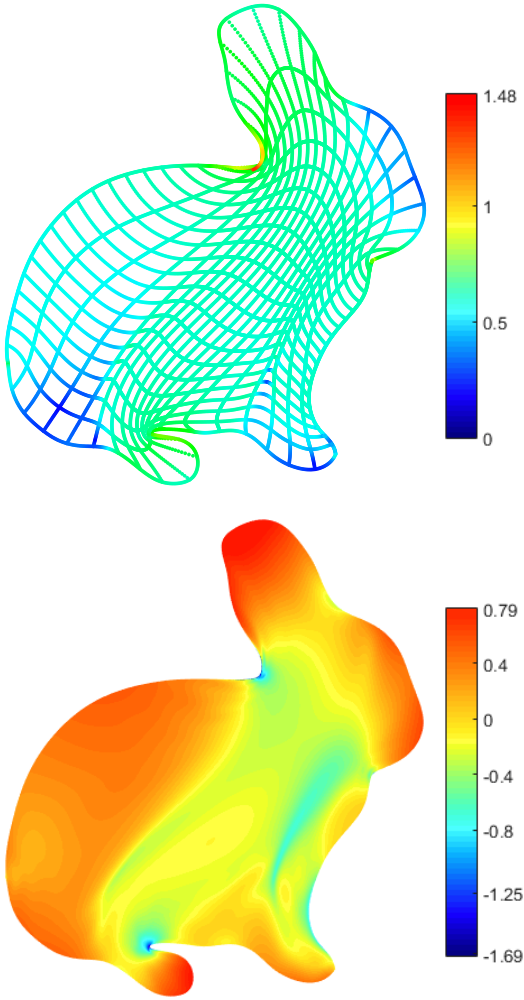}
    }
    \hspace{0.02\textwidth}
    \subfigure[Ours]{
        \label{RabbitmufareaComparison:Ours}
        \includegraphics[width=0.25\textwidth]{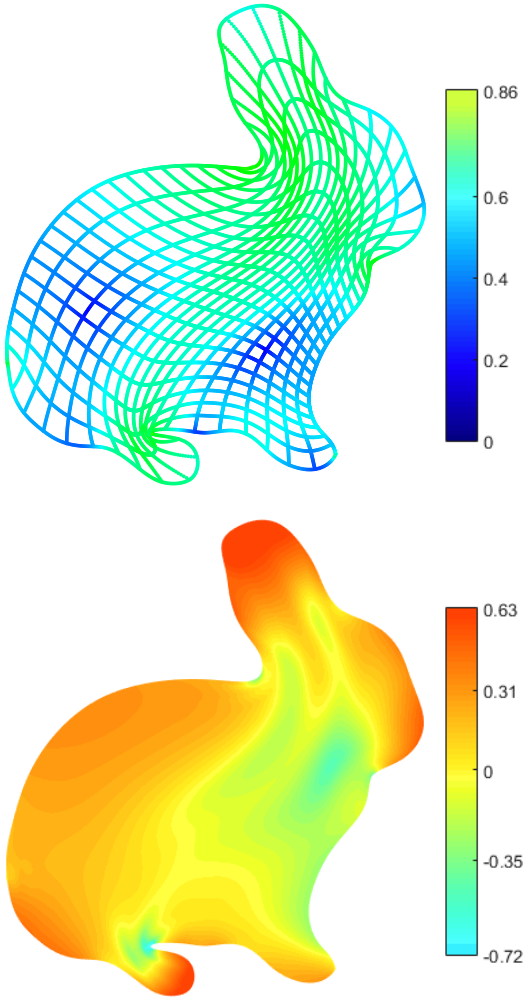}
    }
	\caption{Parameterization the Rabbit model by three methods: \subref{RabbitmufareaComparison:NO} nonlinear optimization, \subref{RabbitmufareaComparison:T-Map} T-map method and \subref{RabbitmufareaComparison:Ours} our method.
 The top and bottom row show the colormaps of $|\mu(f)|$ and $\log_{10}J_s(f)$ respectively.}
    \label{RabbitmufareaComparison}
\end{figure}

\begin{figure}[!htbp]
    \centering
    \subfigure[Nonlinear optimization]{
        \label{ButterflymufareaComparison:NO}
        \includegraphics[width=0.25\textwidth]{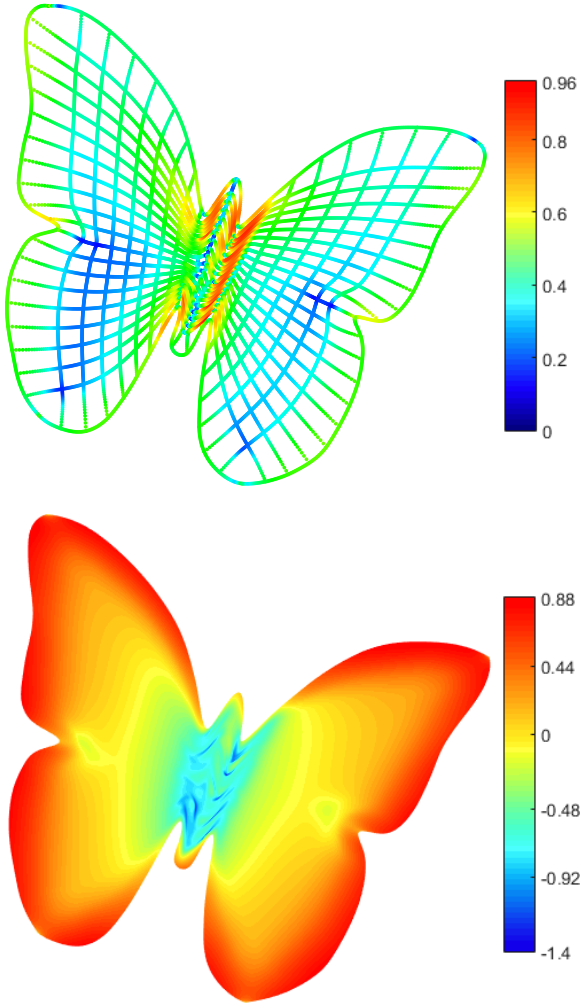}
    }
    \hspace{0.02\textwidth}
    \subfigure[T-map]{
        \label{ButterflymufareaComparison:T-Map}
        \includegraphics[width=0.25\textwidth]{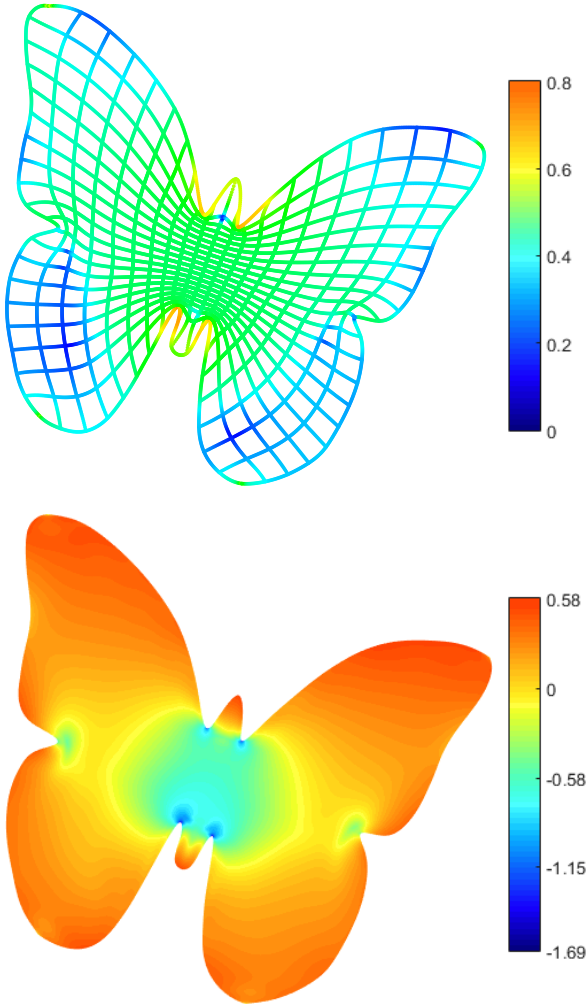}
    }
    \hspace{0.02\textwidth}
    \subfigure[Ours]{
        \label{ButterflymufareaComparison:Ours}
        \includegraphics[width=0.25\textwidth]{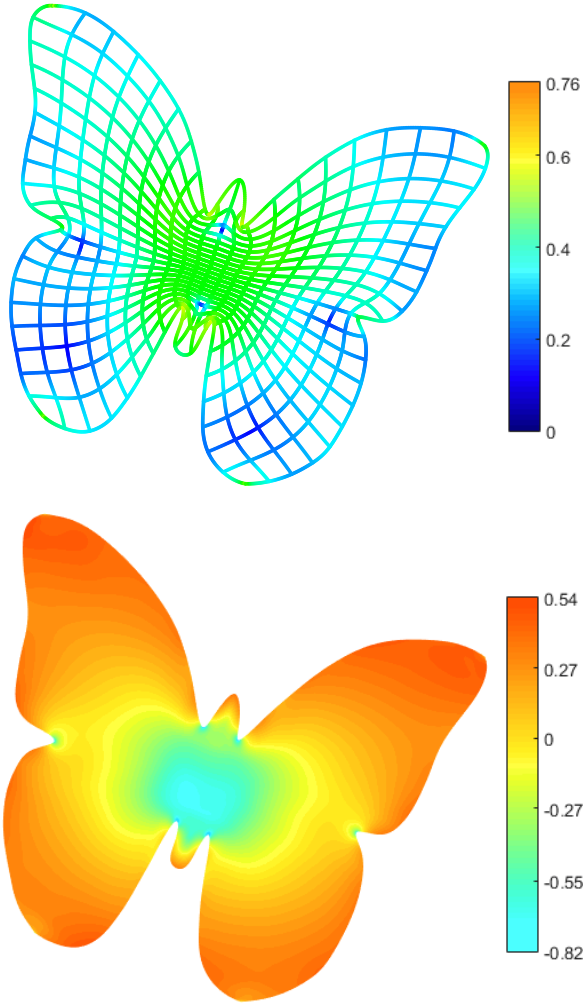}
    }
	\caption{Parameterization the Butterfly model by three methods: \subref{ButterflymufareaComparison:NO} nonlinear optimization, \subref{ButterflymufareaComparison:T-Map} T-map method and \subref{ButterflymufareaComparison:Ours} our method.
 The top and bottom row show the colormaps of $|\mu(f)|$ and $\log_{10}J_s(f)$ respectively.}
    \label{ButterflymufareaComparison}
\end{figure}

\begin{figure}[!htbp]
    \centering
    \subfigure[Nonlinear optimization]{
        \label{DolphinmufareaComparison:NO}
        \includegraphics[width=0.28\textwidth]{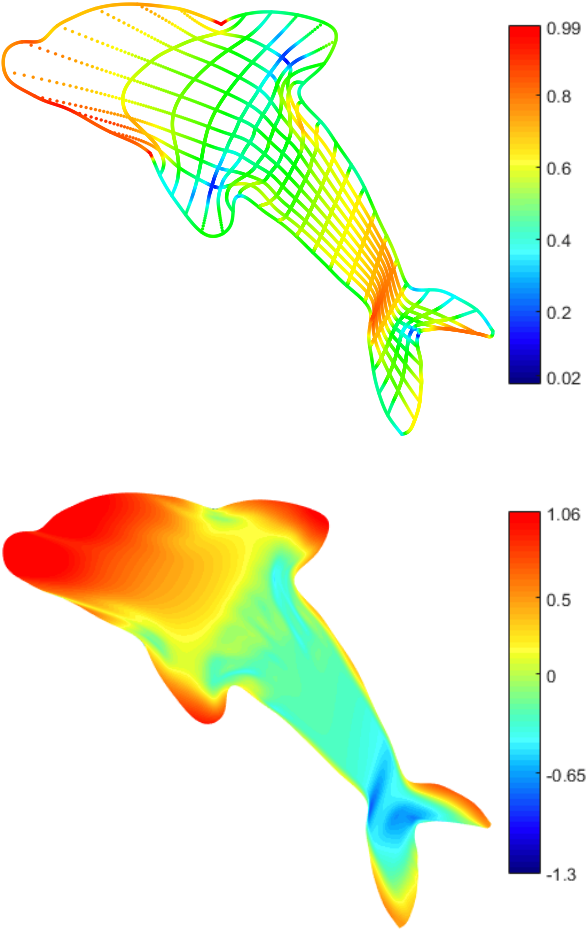}
    }
    \hspace{0.01\textwidth}
    \subfigure[T-map]{
        \label{DolphinmufareaComparison:T-Map}
        \includegraphics[width=0.28\textwidth]{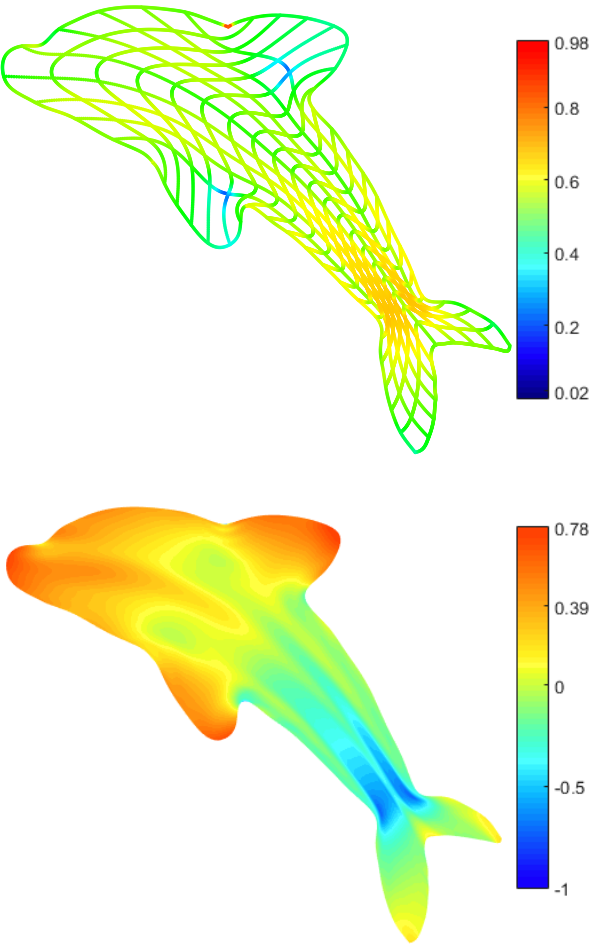}
    }
    \hspace{0.01\textwidth}
    \subfigure[Ours]{
        \label{DolphinmufareaComparison:Ours}
        \includegraphics[width=0.28\textwidth]{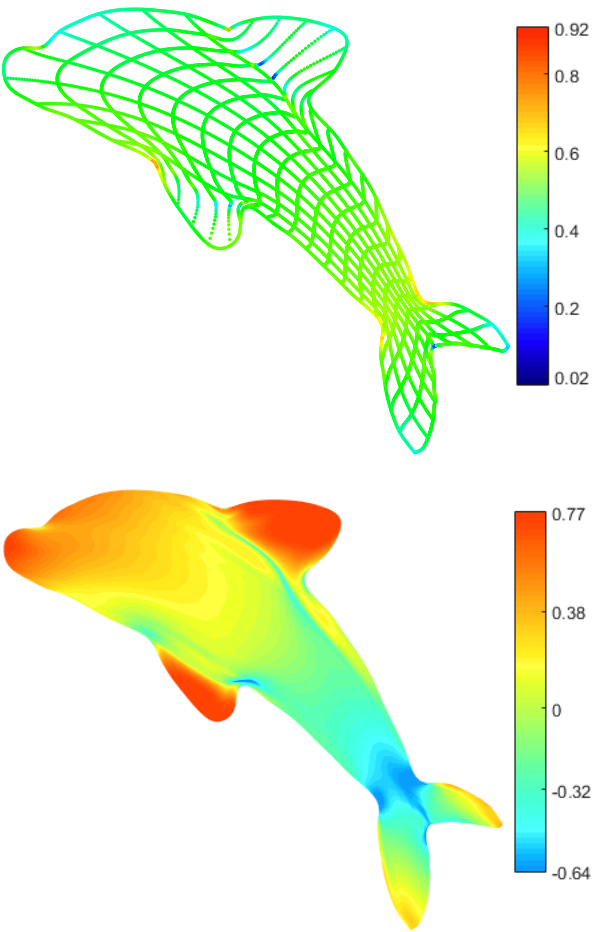}
    }
	\caption{Parameterization the Dolphin model by three methods: \subref{DolphinmufareaComparison:NO} nonlinear optimization, \subref{DolphinmufareaComparison:T-Map} T-map method and \subref{DolphinmufareaComparison:Ours} our method.
 The top and bottom row show the colormaps of $|\mu(f)|$ and $\log_{10}J_s(f)$ respectively.}
    \label{DolphinmufareaComparison}
\end{figure}

From~\autoref{rankcomparision}, we can see that the variational harmonic method and the low-rank spline method are significantly worse than our method in terms of distortion. In Fig.~\ref{RabbitmufareaComparison}, Fig.~\ref{ButterflymufareaComparison}, and Fig.~\ref{DolphinmufareaComparison}, we compare our method with the nonlinear optimization method and T-map method by displaying the colormaps of the Beltrami coefficients $|\mu(f)|$ and the scaled Jacobian $J_s(f)$. It can be seen that our method produces smaller angular distortion in some concave regions, e.g. in the ear of the rabbit, in the root of the wing of the butterfly, and in the body of the dolphin than the other two methods, \ and at the same time, our approach achieves best results in terms of the area distortion among the three methods.

In summary, our method produces much better parameterization results than the other state-of-the-art methods in all these examples. The reason might be as follows. The variational harmonic method and the low-rank spline interpolation method can't guarantee the injectivity in theory. The Teichm{\"u}ller mapping method computes a Teichm{\"u}ller map by solving a nonlinear and non-convex optimization problem. But their method has no convergence guarantee, and thus it may not be able to find the real Teichm{\"u}ller map in some cases. The nonlinear optimization method puts some strict constraints, which could result in no solutions for complex shapes.

\subsubsection{Solving PDEs using IGA}\label{solvingpdes}
In this subsection, we apply our low-rank parameterization together with IGA to solve numerical partial differential equations (PDEs) on different domains. The stability, accuracy and efficiency of the numerical simulation are compared with the nonlinear optimization method and the T-map method.

Consider the following elliptic problem
\begin{equation}\label{elliptic}
\left\{
\begin{aligned}
-\Delta u + u & = f &\text{in}\ &\Omega \\
u|_{\partial \Omega} & = g &\text{on}\ &\partial\Omega
\end{aligned}
\right.
\end{equation}
where $\Omega$ is a Lipschitz continuous domain with boundary $\partial \Omega$, $f,g\in L^2(\Omega)$ are given. The variational form of the problem~(\ref{elliptic}) consists in finding $u\in V=\{u|u\in H^1(\Omega),u|_{\partial \Omega}=g\}$, such that
\begin{equation}\label{variational}
a(u,v) = f(v),\quad \forall v\in H_0^1(\Omega).
\end{equation}
where $$a(u,v) = \int_{\Omega}(\nabla u\cdot\nabla v+uv)\mathrm{d}x\mathrm{d}y,\quad f(v)=\int_{\Omega}fv\mathrm{d}x\mathrm{d}y$$ %\textcolor{red}{dx is not defined!}

Let $u = w+g$, then the problem~(\ref{variational}) is equivalent to find $w\in H_0^1(\Omega)$, such that
\begin{equation}\label{variationalhomogeneous}
a(w,v) = l(v),\quad \forall v\in H_0^1(\Omega).
\end{equation}
where $l(v)=f(v)-a(g,v)$.

In the setting of isogeometric analysis, the domain $\Omega$ is parameterized by a global map $f:\hat{\Omega}\to \Omega$ which is defined in~(\ref{B-spline complexrep}). The isogeometric discretization takes advantages of the given parameterization of the domain $\Omega$. In particular, the discretization space $V_h$ can be chosen as
\begin{equation*}
V_h = \spn\{\phi_{ij}(x,y), i=0,1,\ldots,m, j=0,1,\ldots,n\}
\end{equation*}
with $\phi_{ij}=B_{ij}\circ f^{-1}(x,y)$ and $B_{ij}(x,y)=M_i^p(x)N_j^q(y)$.

The finite-dimensional space $V_h$ is now used for the Galerkin discretization of the variational formulation~(\ref{variationalhomogeneous}), which consists in finding $w_h\in V_h$, such that
\begin{equation}
a(w_h,v_h)=l(v_h),\quad \forall v_h\in V_h.
\end{equation}

We solve the above elliptic problem over two domain examples (the Rabbit-shaped domain and the Butterfly-shaped domain) to show the numerical advantages of our low-rank parameterization method.

\medskip

\begin{figure}[!htbp]
    \centering
    \subfigure[Nonlinear optimization]{
        \label{IGARabbitError:NO}
        \includegraphics[width=0.30\textwidth]{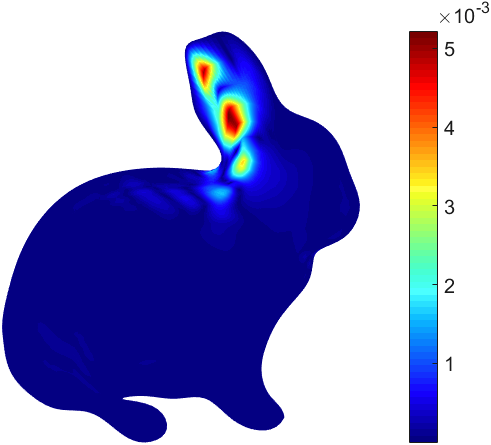}
    }
    \hspace{-0.02\textwidth}
    \subfigure[T-map]{
        \label{IGARabbitError:T-Map}
        \includegraphics[width=0.30\textwidth]{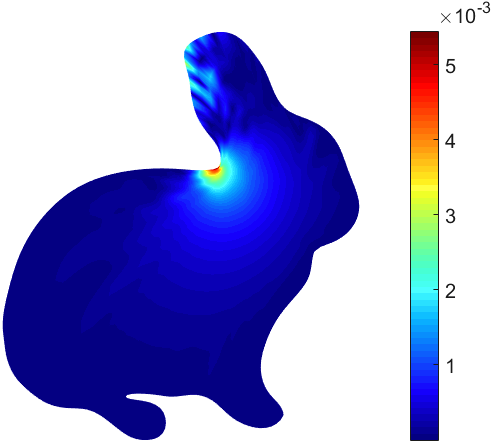}
    }
    \hspace{-0.02\textwidth}
    \subfigure[Ours]{
        \label{IGARabbitError:Ours}
        \includegraphics[width=0.30\textwidth]{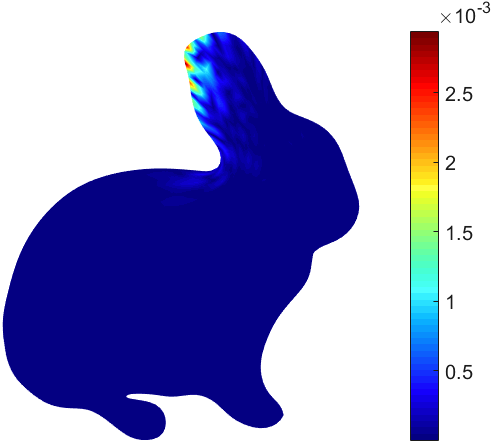}
    }
	\caption{Rabbit-shaped domain: The numerical errors of the solutions for~\subref{IGARabbitError:NO} nonlinear optimization method,~\subref{IGARabbitError:T-Map} T-map method and~\subref{IGARabbitError:Ours} our method.}
    \label{IGARabbitError}
\end{figure}

\noindent\textbf{Rabbit-shaped domain with different parameterizations}
In this example, we solve the elliptic problem over the Rabbit-shaped domain, where $u$ has an exact solution $1.0/((x-0.5)^2+(y-0.5)^2+0.02)$. The parameterization results of this domain are shown in Fig.~\ref{RabbitInjectivityComparison}. The degrees of freedom ($DOF$) of the basis functions in $V_h$ in this example is $2209$. Fig.~\ref{IGARabbitError:NO},~\ref{IGARabbitError:T-Map} and~\ref{IGARabbitError:Ours} show the numerical errors of the solutions for the nonlinear optimization method, the T-map method and our method respectively, and Table~\ref{igarabbitcomparison} summaries the condition numbers of the stiffness matrices, $L_2$ errors and the assembling time for these three methods. We can see that our method produces smaller condition numbers and errors than the other two methods. At the same time, owing to the low-rank property, our method can accelerate the assembly process of the matrices in IGA by using the low-rank assembly strategy presented in~\citep{mantzaflaris2014matrix,mantzaflaris2016low}.

\begin{table}[!htbp]
	\centering
	\scriptsize
    \renewcommand{\arraystretch}{1.2}
    \def\temptablewidth{0.38\textwidth}
	\begin{tabular}{|c|c|c|c|}\hline
    Method & Condition number &$L_2$ error& Assembling time (s)\\
    \cline{1-4}
        Nonlinear optimization &4857&0.0002428&72.66\\
        \hline
        T-map & 7868 &0.000669&71.96\\
		\hline
        Ours& 1627 &0.000011&31.48\\
        \hline
	\end{tabular}
	\caption{Comparisons of the condition numbers of the stiffness matrices, $L_2$ errors and the computational time (in seconds) for assembling the stiffness matrices for the Rabbit-shaped domain.}
    \label{igarabbitcomparison}
\end{table}

\medskip

\begin{figure}[!htbp]
    \centering
    \subfigure[Nonlinear optimization]{
        \label{IGAButterflyError:NO}
        \includegraphics[width=0.30\textwidth]{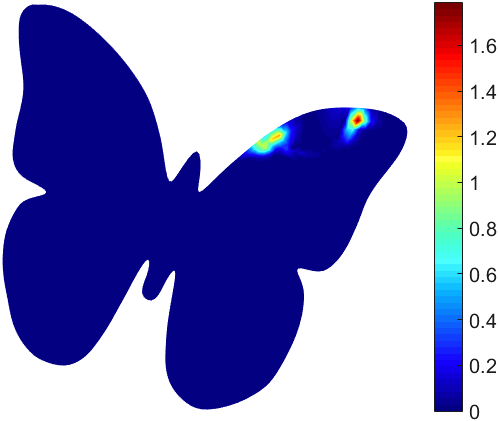}
    }
    \hspace{-0.02\textwidth}
    \subfigure[T-map]{
        \label{IGAButterflyError:T-Map}
        \includegraphics[width=0.30\textwidth]{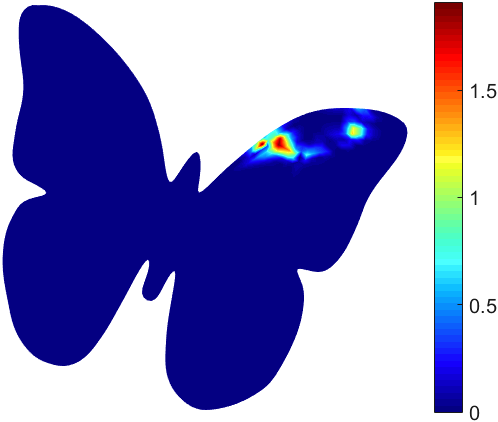}
    }
    \hspace{-0.02\textwidth}
    \subfigure[Ours]{
        \label{IGAButterflyError:Ours}
        \includegraphics[width=0.30\textwidth]{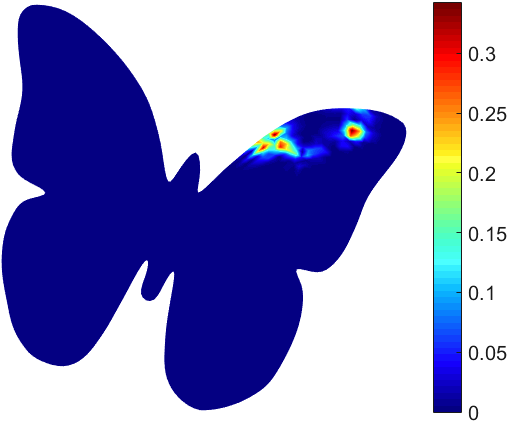}
    }
	\caption{Butterfly-shaped domain: The numerical errors of the solutions for~\subref{IGAButterflyError:NO} nonlinear optimization method,~\subref{IGAButterflyError:T-Map} T-map method and~\subref{IGAButterflyError:Ours} our method.}
    \label{IGAButterflyError}
\end{figure}

\noindent\textbf{Butterfly-shaped domain with different parameterizations}
We consider the elliptic problem~(\ref{elliptic}) over another domain---the Butterfly-shaped domain, where the parameterization results of the nonlinear optimization method, T-map method and our method are shown in Fig.~\ref{ButterflyInjectivityComparison}. The $DOF$ of the basis functions in $V_h$ and the exact solution in this example are $8281$ and $\tanh((0.25-\sqrt{(x-0.5)^2+(y-0.5)^2})/0.03)$ respectively. Fig.~\ref{IGAButterflyError:NO},~\ref{IGAButterflyError:T-Map},~\ref{IGAButterflyError:Ours} show the numerical errors of the solutions and Table~\ref{igabutterflycomparison} lists the condition numbers of the stiffness matrices, $L_2$ errors and the assembling time for these three methods. Again we can see that our method produces smaller condition numbers and errors than the other two methods, and at the same time, our method can accelerate the assembly process of the stiffness matrices in IGA.

\begin{table}[!htbp]
	\centering
	\scriptsize
    \renewcommand{\arraystretch}{1.2}
    \def\temptablewidth{0.38\textwidth}
	\begin{tabular}{|c|c|c|c|}\hline
    Method & Condition number &$L_2$ error& Assembling time (s)\\
    \cline{1-4}
        Nonlinear optimization &43105&0.00422&391.89\\
        \hline
        T-map & 9074 &0.001862&390.78\\
		\hline
        Ours& 3834 &0.000652&170.19\\
        \hline
	\end{tabular}
	\caption{Comparisons of the condition numbers of the stiffness matrices, $L_2$ errors and the computational time (in seconds) for assembling matrices for the Butterfly-shaped domain.}
    \label{igabutterflycomparison}
\end{table}

\section{Conclusions and future work}
\label{sec:conclusion}
Parameterization of computational domains and efficiently assembling the mass and stiffness matrices are two essential steps in isogeometric analysis applications. In this paper, using low-rank tensor approximation technique, we propose a low-rank representation scheme for domain parametrization based on quasi-conformal mapping. The problem is formulated as a non-linear and non-convex optimization problem which minimizes the angular distortion and the rank of the map while ensuring the bijectivity of the map. The optimization problem is then converted into two quadratic optimization problems which are solved alternatively. Several experimental examples show that our approach can produce a low-rank and low-distortion parameterization which is superior to other state-of-the-art methods. Numerical examples of our parameterization method together with IGA in solving numerical PDEs also demonstrate some numerical advantages of our method than previous approaches.

Regarding the future work, extending our work to three-dimensional volumetric parametrization is worthy of further research. However, the parametrization problem in three dimensional case is much harder since there is no analogous complex structure in three-dimensional space.

%!Mode:: "TeX:UTF-8"

\section*{Acknowledgement}
\label{sec:acknowledgement}
This work is supported by the NSF of China (No. 11571338, 11626253) and by the Fundamental Research Funds for the Central Universities (WK0010000051).

\end{document}